\documentclass[showpacs,showkeys,twocolumn]{revtex4-1}
\usepackage{bm,graphicx,subfigure,color}

\newcommand*\blue{\color{blue}}

\begin{document}
\title{Ground and Low-Lying Collective States of Rotating Three-Boson System}
\author{Mohd. Imran}\email{alimran5ab@gmail.com}
\author{M. A. H. Ahsan}\email{mahsan@jmi.ac.in}
\affiliation{Department of Physics, Jamia Millia Islamia 
           (Central University), New Delhi 110025, India.}
\begin{abstract} 
The ground and low-lying collective states of a rotating system of $N=3$ bosons harmonically confined in quasi-two-dimension and interacting via repulsive finite-range Gaussian potential is studied in weakly to moderately interacting regime. The $N$-body Hamiltonian matrix is diagonalized in subspaces of quantized total angular momenta $0\le L \le 4N$ to obtain the ground and low-lying eigenstates. Our numerical results show that breathing modes with $N$-body eigenenergy spacing of $2\hbar\omega_{\perp}$, known to exist in strictly 2D system with zero-range ($\delta$-function) interaction potential, may as well exist in quasi-2D system with finite-range Gaussian interaction potential. To gain an insight into the many-body states, the von Neumann entropy is calculated as a measure of quantum correlation and the conditional probability distribution is  analyzed for the internal structure of the eigenstates. In the rapidly rotating regime the ground state in angular momentum subspaces $L=\frac{q}{2}N\left(N-1\right)$ with $q=2, 4$ is found to exhibit the anticorrelation structure suggesting that it may variationally be described by a Bose-Laughlin like state. We  further observe that the first breathing mode exhibits features similar to the Bose-Laughlin state in having eigenenergy, von Neumann entropy and internal structure independent of interaction for the three-boson system considered here. On the contrary, for eigenstates lying between the Bose-Laughlin like ground state and the first breathing mode, values of eigenenergy, von Neumann entropy and internal structure are found to vary with interaction.
\end{abstract}
\maketitle

\section{Introduction}
\label{intro}
Advances made in past few decades have given experimentalists a decisive control over several parameters of  physical interest of quantum gases such as the density, the effective dimensionality, the inter-particle interaction \cite{ias98,cgj10} and has even made it feasible to confine the desired number of particles as in a microchip trap \cite{lsh03,fz07}.
Following this, considerable effort has been devoted to explore the physics of few-body systems of interacting particles \cite{yl07}, as it promises to provide a bridge between the microscopic and the macroscopic ensembles \cite{bdz08,blu12}.
As a consequence, an increasing number of few-body phenomena are being studied in a variety of systems under different physical situations \cite{bhh02,bh06,zms06,gps08,pm14}.
There are other compelling reasons to consider systems with few particles, for instance, a strongly correlated state of the quantum gas is experimentally accessible only for small systems, such as an analogue of the Laughlin-like state \cite{rbl83,rbl99,mr91,gjl14}. 
Another reason is that few-body systems allow a higher level of control \cite{szl11,hhm10}, to study its role as building blocks of strongly correlated quantum many-body states \cite{blu12}.
\\
\indent
Recent studies have demonstrated that the breathing mode \cite{pr97,cbr02} is ideally suited to be employed as a diagnostic tool for probing the ground and the excited states of quantum gases in trapped atomic vapors \cite{moa13,abm14} paving the way for a novel kind of spectroscopy of ultracold trapped gases.
For classical system of interacting particles confined in 2D by external potentials, breathing mode 
has been studied earlier \cite{sp95,hfl08,obm09}.
Quantum few-body systems realized in lower dimensions with high-precision control over its physical parameters serves to motivate the study of breathing mode dynamics \cite{bhh02,abm14,opl10,tmh13,im15}. 
An understanding of few-body systems may provide an insight into the beyond mean-field physics of macroscopic ensembles.
\\
\indent
In this work, we present an exact diagonalization study of ground and excited states including the breathing modes in a system of three spinless bosons, harmonically confined and interacting via finite-range Gaussian repulsive potential.
The results obtained through exact diagonalization method attribute a $2\hbar \omega_{\perp}$ spacing that exists between specific collective excitations, referred to as {\it breathing modes} of the system in 2D harmonic trap.
In particular, the paper focuses on the case of rapidly rotating bosons (in quantum Hall regime) to explore the quantum correlation as well as internal structures of the many-body ground and collective excited states.
Systems in fractional quantum Hall regime characterized by finite number of particles or high angular velocity approaching the centrifugal limit of the confining harmonic trap are of special interest \cite{gjl14,rps02,zsg14} as the ground state of such a system is the well studied Bose-Laughlin state \cite{cw99,wg00,coo08}.
Anticipating experiments with few Bose atoms, the purpose of this study is to further extend theoretical understanding in this direction. 
\\
\indent
The paper is organized as follows. 
Section~\ref{sec:model} provides a brief description of the model Hamiltonian for a rotating Bose system with finite-range Gaussian interaction potential and confined in quasi-two-dimensional (quasi-2D) harmonic trap. 
In section~\ref{sec:results}, we present the results for a system of three ultra-cold Bose atoms to investigate the physics of breathing modes. 
For weakly to moderately interacting regime, the low-energy eigenspectra is presented to examine the ground and low-lying collective excitations in rapidly rotating regime. 
In order to gain further insight into the few-body quantum states, we obtain the von Neumann entropy as a measure of quantum correlation and the conditional probability distribution to analyze the internal structure (spatial correlation). 
Section~\ref{sec:conc} presents summary and outline the conclusions of the present work.
\section{Theoretical Model}
\label{sec:model}
\subsection{The system and the Hamiltonian}
We consider a system of interacting spinless bosons, harmonically confined and subjected to an externally impressed rotation with angular velocity ${\widetilde{\bm{\Omega}}}\equiv \widetilde{\Omega} \hat{e}_{z}$ about the $z$-axis. 
We assume a stiff confinement of the harmonic trap along the axis of rotation so that the axial energy-level spacing far exceeds the radial energy-level spacing and other energy scales like the interaction energy and the rotational energy, yielding an effectively quasi-2D system with $x$-$y$ rotational symmetry.
Choosing $\hbar \omega_{\perp}$ and $ a_{\perp} = \sqrt{\hbar/{M \omega_{\perp}}}$ as units of energy and length respectively, our system of $N$ spinless bosons each with mass $M$ and radial confining frequency $\omega_{\perp}$ is described in the co-rotating frame by the Hamiltonian $H^{rot} = H^{lab}-\Omega L$, where
\begin{equation}
H^{lab} = \sum_{j=1}^{N} \left[-\frac{1}{2} \bm{\nabla}^{2}_{j} + \frac{1}{2} {\bf r}_{j}^{2} \right] + \frac{1}{2} \sum_{i\neq j}^{N} U \left(\vert {\bf r}_{i}-{\bf r}_{j}\vert\right)
\label{mbh}
\end{equation} 
Here $\Omega = \widetilde{\Omega}/{\omega_{\perp}}$ $(\leq 1)$ is the dimensionless angular velocity and $L$ (scaled by $\hbar$) is the $z$ projection of the total angular momentum operator.
The first two terms in the Hamiltonian~(\ref{mbh}) correspond to the kinetic and potential energies. 
The third term $U\left(|{\bf r}_{i}-{\bf r}_{j}|\right)$ arises from the two-body interaction assumed to be Gaussian in particle-particle separation~\cite{im15,iasl15}
\begin{equation}
U \left(\vert{\bf r}_{i}-{\bf r}_{j}\vert\right) = \frac{\mbox{g}_{2}} {2\pi{\sigma_{\perp}^{2}}}
\exp{\left[ -\frac{\left({\bf r}_{i}-{\bf r}_{j}\right)^{2}}{2\sigma_{\perp}^{2}} \right]} 
\label{gip}
\end{equation}
with $\sigma_{\perp}$ (scaled by $a_{\perp}$) being the effective range of two-body interaction.
The dimensionless parameter $\mbox{g}_{2}=4\pi {a_{s}}/{a_{\perp}}$ measures the strength of interaction where $a_{s}$ is the $s$-wave scattering length for low-energy particle-particle collision.
\\
\indent
Recent experimental advancements in atomic physics have made it possible to tune $a_{s}$ in ultra-cold atomic vapors using Feshbach resonance \cite{ias98,cgj10}. Accordingly, in a theoretical study, one can vary $a_{s}$ to achieve the desired value of interaction parameter $\mbox{g}_{2}$ relevant to the trapped model system. 
We take the scattering length to be positive $\left(a_{s}>0\right)$ so that the effective interaction is repulsive. The finite-range Gaussian interaction potential in Eq.~(\ref{gip}) is expandable within a finite number of single-particle basis functions and hence computationally more feasible \cite{dka13,cfa09}. 
In the limit $ \sigma_{\perp}\rightarrow 0$, the normalized Gaussian potential in Eq.~(\ref{gip}) smoothly reduces to the zero-range contact potential $\mbox{g}_{2}\delta\left({\bf r}_{i}-{\bf r}_{j}\right)$ which has widely been used in earlier studies \cite{dgp99}.
\\
\indent
The system described by the Hamiltonian in Eq.~(\ref{mbh}) has cylindrical symmetry with respect to the $z$-axis which implies that the $z$-projection of the total angular momentum is conserved {\it i.e.} $L$ is a good quantum number. 
To obtain the eigenenergies and the corresponding eigenstates of the $N$ boson system, we employ exact diagonalization of the Hamiltonian matrix in different subspaces of $L$ with inclusion of lower as well as higher Landau levels in constructing the $N$-body basis states.
\\
\indent
In Rayleigh-Ritz scheme \cite{gok88} employed here, the $N$-body variational wavefunction $\Psi\left({\bf r}_{1},{\bf r}_{2},\dots,{\bf r}_{N}\right)$ is constructed as linear combination of the symmetrized products $\left\{\Phi_{\bm{\nu}} \left({\bf r}_{1},{\bf r}_{2},\dots , {\bf r}_{N}\right) \right\}$ of a finite number of single-particle basis functions $\left\{{u}_{n,m,n_{z}}\left({\bf r}\right)\right\}$, chosen to be the eigenfunctions of the non-interacting single-particle Hamiltonian
\begin{equation}
H_{sp} = \frac{1}{2} \left(-{\bf\nabla}^{2}_{\perp} + r_{\perp}^{2}\right) - {\Omega} \, {\ell_{z}} + \frac{1}{2} \left(-{\bf\nabla}^{2}_{z} + \lambda_{z}^{2} z^{2}\right),
\label{sph2}
\end{equation}
identified as the quasi-2D harmonic oscillator Hamiltonian in a rotating frame with $\ell_{z}$ being the single-particle angular momentum. 
The eigensolutions of ${H}_{sp} \, {u}_{n,m,n_{z}}\left({\bf r}\right)={\epsilon}_{n,m,n_{z}}{u}_{n,m,n_{z}} \left({\bf r}\right)$, in dimensionless form, are known to be:
\begin{eqnarray} 
&&{\epsilon}_{n,m,n_{z}} = \left(n+1-m{\Omega}\right) + {\lambda}_{z}\left(n_{z}+{1}/{2}\right)\nonumber \\
&&{u}_{n,m,n_{z}} \left({\bf r}\right) =
\sqrt{\frac{\left(\frac{1}{2}\left\{ n-|m| \right\} \right)! \, \sqrt{{\lambda}_{z}/\pi^{3}}}{\left(\frac{1}{2}\left\{ n+|m| \right\} \right)! \ 2^{n_{z}} \ n_{z}!}}\ e^{-\left({r_{\perp}^{2}+ {\lambda}_{z} z^{2}}\right)/2} \nonumber \\
&&~~~~~~~~~~~~~~~~ \times e^{im\phi } \ {r^{|m|}_{\perp }}\, 
L^{|m|}_{\frac{1}{2}(n-|m|)} \left(r_{\perp }^{2}\right) H_{n_{z}}\left({\lambda}_{z}z^{2}\right)
\label{sps1}
\end{eqnarray}
where $n=2n_{r}+|m|$ with $n_{r}=0,1,2,\cdots$ and $m=0,\pm 1,\pm 2, \cdots$.
Here $L^{|m|}_{\frac{1}{2}(n-|m|)}\left(r_{\perp }^{2}\right)$ is the associated Laguerre polynomial and $H_{n_{z}}\left({\lambda}_{z}z^{2}\right)$ is the Hermite polynomial. 
Also $n_{r} \equiv \frac{1}{2}(n-|m|)$ is the radial quantum number and $m$ is the single-particle angular momentum quantum number. 
The system here has been assumed to be quasi-2D since there is practically no excitation along the relatively stiffer $z$-axis and we, therefore, set $n_{z}=0$ in Eq.~(\ref{sps1}) implying that all the particles occupy only the lowest-energy state ${u}_{0}(z)=({\lambda}_{z}/\pi)^{1/4}\ e^{-{{\lambda}_{z} z^{2}}/2}$ of $z$ co-ordinate degree of freedom.
Therefore Eq.~(\ref{sps1}) can be written as
\begin{eqnarray} 
&&\epsilon_{n,m}=\left(n+1-m {\Omega}\right)+ {\lambda_{z}}/{2}~,~~\mbox{with}~n=2n_{r}+|m| \nonumber \\
&&{u}_{n,m} \left({\bf r}\right)=
\sqrt{\frac{\left(\frac{1}{2}\left\{ n-|m| \right\} \right)!}{\left(\frac{1}{2}\left\{ n+|m| \right\} \right)!}\sqrt{\frac{{\lambda}_{z}}{\pi^{3}}}}\ e^{-\left({r_{\perp}^{2}+ {\lambda}_{z} z^{2}}\right)/2}
\nonumber \\
&&~~~~~~~~~~~~~~ \times e^{im\phi }\ {r_{\perp }^{|m|}}\ 
L^{|m|}_{\frac{1}{2}(n-|m|)} \left(r_{\perp }^{2}\right).
\label{sps2}
\end{eqnarray}
Restricting to $n_{r}=0$ and taking $m \geq 0$ in the above equation corresponds to the LLL approximation. 
Taking $n_{r} \geq 0$ and allowing $m$ to take positive as well as negative values corresponds to going beyond LLLs \cite{ahs01,lhc01}.
The $N$-body variational wavefunction is
\begin{equation}
\Psi\left({\bf r}_{1},{\bf r}_{2},\dots,{\bf r}_{{N}}\right) = \sum_{\bm{\nu}} {{C}_{\bm{\nu}}} \, \Phi_{\bm{\nu}} \left({\bf r}_{1},{\bf r}_{2},\dots,{\bf r}_{N} \right)
\label{nbf}
\end{equation}
where $\left\{ C_{\bm{\nu}} \right\}$ are the variational parameters.
The many-body index $\bm{\nu}\equiv \left(\nu_{\bf 0},\nu_{\bf 1},\dots,\nu_{\bf j},\dots, \nu_{\bf k}\right)$ labelling the many-body basis function $\Phi_{\bm{\nu}} \left({\bf r}_{1},{\bf r}_{2},\dots, {\bf r}_{N} \right)$ stands for a set of single-particle quantum numbers $\left\{ {\bf j} \equiv \left(n,m\right)\right\}$ and their respective occupancies $\left\{ \nu_{\bf j}\right\}$.
In the present work we employ beyond lowest-Landau level approximation, constructing many-body basis functions $\left\{\Phi_{\bm{\nu }}\right\}$ using the single-particle basis ${u}_{n,m}\left({\bf r}\right)$ with radial quantum number $n_{r}=\frac{1}{2} \left( n-|m|\right)\geq 0 $ and angular momentum quantum number $|m|\geq 0$.
In the second-quantized notation, the Bose field operator can be expanded in terms of single-particle basis states as $\hat{\psi} \left({\bf r}\right) = \sum_{\bf j}\hat{b}_{\bf j} u_{\bf j} \left({\bf r}\right)$.
In occupation-number representation, the $N$-body basis function $\vert \Phi_{\bm{\nu}} \rangle$ is written  in second-quantized form as:  
\begin{equation}
\vert \Phi_{\bm{\nu}} \rangle \equiv  \prod_{\bf j=0}^{\bf k}
\frac{1}{\sqrt{\nu_{\bf j}!}}{\left(\hat{b}^{\dagger}_{\bf j}\right)}^{\nu_{\bf j}} \vert \mathrm{vac} \rangle 
\equiv  \vert \nu_{\bf 0}\ \nu_{\bf 1}\cdots \nu_{\bf j} \cdots \nu_{\bf k}\rangle 
\label{nbb}
\end{equation}
with
$\sum_{{\bf j}={\bf 0}}^{\bf k}\nu_{\bf j}={{N}}$ and
$\sum_{{\bf j}={\bf 0}}^{\bf k}m_{\bf j}\nu_{\bf j}={L}$ where ${\bf j}=(n_{\bf j},m_{\bf j})$.
With these constraints, only the most important Fock states from the full basis with a given $L$ (the active Fock space) are included.
Once the active Fock states are constructed as in (\ref{nbb}), we diagonalize the Hamiltonian matrix.
Details of the diagonalization scheme and beyond lowest Landau level approximation employed here, has been presented in Ref.~\cite{ahs01}.
\subsection{Diagonalization of the Hamiltonian}
Once the active Fock states are constructed, we calculate the matrix elements and subsequently diagonalize the  Hamiltonian matrix.
For $N=3$ bosons, we have carried out calculations for all the total angular momentum states in the regime $0\leq L \leq 4N$. Diagonalization of the Hamiltonian matrix is performed for each of the subspaces of $L$ separately. 
We have set $n_{z}=0$ in the single-particle basis function $u_{n_{z}}\left(z\right)$ since there is practically no excitation along the relatively stiffer $z$-axis. For a given subspace $L$, the single-particle basis $ u_{n,m}\left(r_{\perp },\phi\right)$ spanning the 2D $xy$ plane is chosen as follows. 
\\
\indent
It is convenient to define  $\ell_{z} \equiv\left[L/N\right]$ where for real $x$ the symbol $\left[x\right]$ denotes the greatest integer less than or equal to $x$. The single-particle angular momentum for the basis functions is now chosen to be: $m=\ell_{z}-n_{b},\ \ell_{z}-n_{b}+1,\ \cdots \ell_{z}+n_{b}-1,\ \ell_{z}+n_{b}$, where $n_{b}$ is some positive integer that we have chosen to be 3, 4 or more depending on the strength of the interaction and the computational resources available ($n_{b}$ is a kind of the size of the single-particle basis chosen for calculation for a given value of $L$). 
The single-particle basis functions thus chosen are used to construct the variational trial function $\Psi =\sum_{\bm{\nu }}C_{\bm{\nu}}~\Phi_{\bm{\nu}}$ of the system for the given value of total angular momentum $L$.
\\
\indent
Since, the system is subjected to an externally impressed rotation about $z$-axis with angular velocity $\Omega$, we diagonalize the many-body Hamiltonian $H^{lab}$ in given subspaces of $L$ to obtain the energy in the corotating frame $E^{rot}\left(L,\Omega\right)=E^{lab}\left(L\right)-\Omega L$. 
This can be seen as the minimization of $E^{lab}(L)$ subject to the constraint that the system has angular momentum expectation value $L$ and the angular velocity $\Omega$ is then the corresponding Lagrange multiplier.
Fixing $L$ fixes $\Omega$ and accordingly we mention $L(\Omega)$ instead of rotational angular velocity $\Omega$, in all the tables and figures in the manuscript.
\section{Results and Discussion}
\label{sec:results}
The results presented here are for a system of $N=3$ Rubidium-$87$ Bose atoms confined in a quasi-2D harmonic trap, interacting via repulsive finite-range Gaussian potential. 
The confining trap frequency is taken to be  ${\omega}_{\perp}=2\pi \times 220$ Hz with the $z$-asymmetry parameter $\lambda_{z}\equiv {\omega_{z}}/{\omega_{\perp}}=\sqrt{8}$ so that the system has small extension $a_{z}=\sqrt{\hbar/M\omega_{z}}$ in the $z$-direction and the dynamics along this axis is assumed to be completely frozen. 
The parameters of the two-body interaction potential (\ref{gip}) have been chosen $\sigma_{\perp}=0.1$ and the $s$-wave scattering length in weakly to moderately interacting regime as $a_{s}=10a_{0}$, $100a_{0}$, $1000a_{0}$ with $a_{0}=0.05292~nm$ being the Bohr atomic radius.
\begin{figure}[!htb]
\centering
\includegraphics[width=0.8\linewidth]{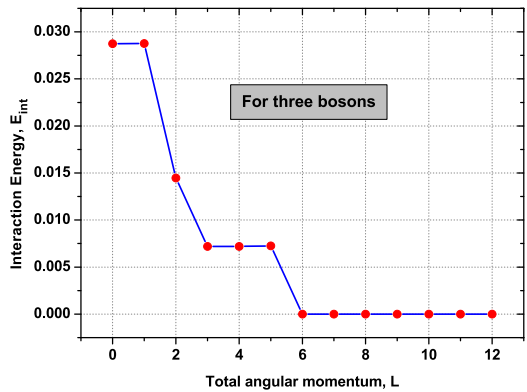}
\caption{\label{fig:n3el} The variation of interaction energy $E_{int}$ contribution of the lowest eigenstates with quantized total angular momentum $L$ ({\it i.e.} yrast line), for the system of $N=3$ bosons interacting via Gaussian potential~(\ref{gip}) with fixed value of interaction range $\sigma_{\perp}=0.1$. For a given $L$, the interaction energy is obtained as $E_{int}(L,\mbox{g}_{2})=E(L,\mbox{g}_{2}=0.09151)-E(L,\mbox{g}_{2}=0.0)$, where $E(L,\mbox{g}_{2})$ is the total energy of the system including the one-body as well as two-body energy terms such as kinetic, potential, rotational and interaction energies, in units of $\hbar \omega_{\perp}$.}
\end{figure}
\begin{table*}[!htb]
\caption{\label{tab:n3a}(Color online) The eigenenergy $E(L_{i})$ of the $L_{i}$ states for $N=3$ bosons in the total angular momentum regime $0 \leq L \leq 12$ with interaction parameters $\mbox{g}_{2}=0.09151$ and $\sigma_{\perp}=0.1$ of the Gaussian potential~(\ref{gip}). The eigenenergy (in units of $\hbar \omega_{\perp}$) of first ten low-lying eigenstates is shown here. The $L_{1}$ states corresponding to $i=1$ (first row in the table) are the yrast states (or ground modes) and $L_{i}$ states $0_{2}$, $1_{2}$, $2_{3}$, $3_{4}$, $4_{5}$, $5_{5}$, $6_{7}$, $7_{7}$, $8_{7}$, $9_{9}$, ${10}_{9}$, ${11}_{8}$, ${12}_{9}$ are the first breathing modes. The states $0_{4}$, $1_{6}$ and $2_{8}$ are the second breathing modes.}
\begin{ruledtabular}
\begin{tabular}{cccccccccccccc}
${i}$ & $L=0$ & $L=1$ & $L=2$ & $L=3$ & $L=4$ & $L=5$ & $L=6$ & $L=7$ & $L=8$ & $L=9$ & $L=10$ & $L=11$ & $L=12$ \\ \hline
1 & {\blue 7.2714} & {\blue 8.2714} & {\blue 9.2571} & {\blue 10.2498} & {\blue 11.2498} & {\blue 12.2507} & {\blue 13.2426} & {\blue 14.2426} & {\blue 15.2426} & {\blue 16.2426} & {\blue 17.2426} & {\blue 18.2426} & {\blue 19.2426} \\
2 & {\blue 9.2713} & {\blue 10.2571} & 9.2715 & 10.2571 & 11.2535 &	12.2522 & 13.2503 & 14.2509 & 15.2498 & 16.2426 & 17.2426 & 18.2471 & 19.2426 \\
3 & 9.2715 & 10.2640 & {\blue 11.2506} & 10.2714 & 11.2571 & 12.2559 & 13.2522 & 14.2522 & 15.2515 & 16.2506 & 17.2491 & 18.2493 & 19.2503 \\
4 & {\blue 11.2501} & 10.2713 & 11.2571 & {\blue 12.2498} & 11.2715 & 12.2713 & 13.2528 & 14.2523 & 15.2522 & 16.2512 & 17.2514 & 18.2516 & 19.2512 \\
5 & 11.2571 & 10.2715 & 11.2614 & 12.2512 & {\blue 13.2481} & {\blue 14.2436} & 13.2565 & 14.2556 & 15.2544 & 16.2522 & 17.2521 & 18.2522 & 19.2518 \\
6 & 11.2587 & {\blue 12.2464} & 11.2686 & 12.2540 & 13.2510 & 14.2506 & 13.2714 & 14.2711 & 15.2706 & 16.2524 & 17.2523 & 18.2536 & 19.2522 \\
7 & 11.2641 & 12.2504 & 11.2715 & 12.2571 & 13.2513 & 14.2515 & {\blue 15.2426} & {\blue 16.2426} & {\blue 17.2434} & 16.2554 & 17.2545 & 18.2697 & 19.2542 \\
8 & 11.2669 & 12.2513 & {\blue 13.2454} & 12.2586 & 13.2523 & 14.2519 & 15.2451 & 16.2432 & 17.2455 & 16.2711 & 17.2705 & {\blue 20.2432} & 19.2703 \\
9 & 11.2713 & 12.2581 & 13.2507 & 12.2626 & 13.2553 & 14.2530 & 15.2498 & 16.2479 & 17.2479 & {\blue 18.2426} & {\blue 19.2426} & 20.2451 & {\blue 21.2426} \\
10 & 11.2715 & 12.2595 & 13.2524 & 12.2699 & 13.2582 & 14.2555 & 15.2512 & 16.2505 & 17.2506 & 18.2430 & 19.2431 & 20.2475	& 21.2429\\
\end{tabular}
\end{ruledtabular}
\end{table*}
The corresponding values of the dimensionless interaction parameter $\mbox{g}_{2}={4\pi a_{s}}/{a_{\perp}}$ turn out to be $0.009151$, $0.09151$ and $0.9151$.
The few-body eigenstates are obtained by diagonalizing the Hamiltonian matrix in each of the subspaces of total angular momentum $0 \leq L\leq 4N$ corresponding to slowly to rapidly rotating regime.
The low-energy eigenspectra for $N=3$ bosons with interaction parameter $\mbox{g}_{2}=0.09151$ is presented in Table~\ref{tab:n3a}, exhibiting how the ground and excited state energies evolve as $L$ is increased. 
Eigenstates in a given subspace of total angular momentum (columns in Table~\ref{tab:n3a}) constitute a $L$ series (or band). The $i$th eigenstate of the $L$ series is denoted by $L_{i}$ and the corresponding eigenenergy by $E\left(L_{i}\right)$.
The lowest energy eigenstate (corresponding to $i=1$) with angular momentum $L_{1}$ is referred to as the yrast state of the $L$ series \cite{mot99}.
The yrast line is drawn by plotting the interaction energy contribution of the lowest-energy eigenstate for each of the $L$ subspaces \cite{yl10,blo06}.
In Fig.~\ref{fig:n3el}, we present interaction energy (in units of $\hbar \omega_{\perp}$) of the yrast states for total angular momenta $0 \le L \le 4N$.
The red solid circles joined by blue line denotes the yrast line. The initial points of the plateaus at $L=0$, $3$ and $6$ are the stable ground states (other points on the plateaus  correspond to metastable states).
\paragraph*{Breathing modes.}
Pitaevskii and Rosch \cite{pr97} demonstrated that the Hamiltonian of a  harmonically confined 2D system with zero-range ($\delta$-function) interaction potential possesses $SO(2,1)$ symmetry due to the scaling behavior of $\delta$-function interaction potential.
This leads to an eigenenergy spectrum with energy spacing of $2\hbar\omega_{\perp}$ between two adjacent breathing modes, describing pulsation of the system.
The above proposition of breathing modes for the non-rotating case \cite{pr97} has further been generalized to slow rotating \cite{sw01,cs03,cbw03,mkm04} as well as rapidly rotating regime \cite{wat06}.
We draw upon these work \cite{bhh02,abm14} to examine the breathing modes in the following.
\\
\indent
It is observed from Table~\ref{tab:n3a} that for the non-rotating state $L=0$ corresponding to $\Omega =0$, the energy interval $E(0_2)-E(0_1)=9.2713-7.2714 =1.9999$ (in units of $\hbar \omega_{\perp}$). 
The states $0_1$ and $0_2$ are respectively the yrast state and the first breathing mode in the subspace $L=0$. 
It is further observed that for the rotating states $L > 0$ too, the energy interval has a value close to $2$. For example, $E({2}_{3})-E({2}_{1})=1.9935$ for $L=2$, $E({5}_{5})-E({5}_{1})=1.9929$ for $L=5$ and $E({11}_{8})-E({11}_{1})=2.0006$ for $L=11$, all close to $2$ (in units of $\hbar\omega_{\perp}$). 
Thus the breathing modes demonstrated \cite{pr97} to exist in a strictly 2D system with zero-range interaction potential are found to be observed here in a more realistic quasi-2D system with finite-range Gaussian interaction potential~(\ref{gip}). 
With an aim to study the physics of breathing mode in rapidly rotating regime, we focus on $L=6$ and $L=12$ angular momentum states of the rotating three-boson system.

\paragraph*{Breathing modes in rapidly rotating regime.}
Increase in total angular momentum leads to increase in (rotational) kinetic energy and decrease in repulsive interaction energy of the ground state due to centrifugal action which moves the bosons apart making their positions more correlated. 
For the ground state of a rapidly rotating system in 2D with $L=N(N-1)$, the interaction energy reduces to zero and the state is found to be the so called strongly correlated Bose-Laughlin state \cite{cw99,coo08} 
\begin{eqnarray}
&&\Psi_{\frac{q}{2}N\left(N-1\right)}^{BL} \left(\left\{ z_{i}\right\} \right) \propto \prod_{i < j} \left(z_{i}-z_{j}\right)^{q}\ \exp{\left(-\frac{1}{2}\sum_{i=1}^{N}\left|z_{i}\right|^{2}\right)} \nonumber \\ 
&&~\mbox{with}~~ L=\frac{q}{2}N\left(N-1\right)~~\mbox{and}~~ q=2,4,\cdots
\label{bl}
\end{eqnarray}
Here $\left( z_{1},z_{2},\cdots ,z_{N}\right)\equiv \left\{z_{i}\right\}$ with $z_{i}={r_{i}}e^{i\phi_{i}}$ denote the dimensionless co-ordinate of the $i$th boson in the complex plane. 
For $q=2,4,\cdots ,$ the Bose-Laughlin state (\ref{bl}) becomes an exact eigenstate of the interaction potential in Eq.~(\ref{gip}) with limiting value $\sigma_{\perp} \rightarrow 0$ {\it i.e.} the $\delta$-function potential. 
The filling fraction $\nu $, defined (in the thermodynamic limit) as the ratio of the total number of particles to the average number of vortices is given by $\nu =1/q$ for the state $\Psi^{BL}_{\frac{q}{2}N(N-1)} \left(\left\{ z_{i}\right\} \right)$. The exponent $q$, therefore, fixes the filling fraction or equivalently the total angular momentum and hence symmetry of the wavefunction. 
\\
\indent
In the present study of $N=3$ rapidly rotating bosons, we confine ourselves to the first two values $q=2$ and $q=4 $ for which the filling fractions are $\nu=1/2$ and $1/4$ respectively and the corresponding angular momenta are $L=6$ and $L=12$. 
The yrast states $6_{1}$ and $12_{1}$ in Table~\ref{tab:n3a} appear as the $q=2$ and $q=4$ Bose-Laughlin states, respectively.
\\
\indent
We first consider the $L=6$ subspace to examine the ground state and the low-lying excited states, obtained variationally through exact diagonalization. 
In Table~\ref{tab:bql6}, we present the eigenenergy $E(L_{i}=6_{i})$ as well as the von Neumann entropy $S(L_{i}=6_{i})$ of low-lying $L_{i}$ states for three representative values of the repulsive interaction parameter $\mbox{g}_{2}$ in the weakly to moderately interacting regime. 
It is seen from the table that the eigenenergy of the yrast state $6_{1}$ is independent of the interaction parameter and the corresponding wavefunction takes the limiting form of Bose-Laughlin state~(\ref{bl}) with $q=2$ for which the interaction energy is zero. 
It is further observed from the table that the energy of the first breathing mode $6_{7}$ is also independent of the interaction parameter and even as the interaction parameter is varied over several orders of magnitude, the ordinal position of the first breathing mode $6_{7}$ with respect to the Bose-Laughlin state $6_{1}$, in the eigenspectrum, remains unchanged\footnote{It must be noted here that in a given subspace of total angular momentum $L$, the ordinal position of the breathing mode ($6_{7}$ for the present case) in the eigenspectrum depends on the size of the active Fock space chosen in the variational calculation. The ordinal position of the breathing mode remains unchanged on varying the interaction for a fixed active Fock space.}.
\begin{table}[!htb]
\caption{\label{tab:bql6}For $N=3$ rapidly rotating bosons in total angular momentum subspace $L=6$, values of eigenenergy ($E$) and von Neumann entropy ($S$) of the ground state and the low-lying excited states including the first breathing mode, with interaction parameters $\mbox{g}_{2}=0.009151$, $0.09151$, $0.9151$ and range $\sigma_{\perp}=0.1$ of the Gaussian potential~(\ref{gip}). The states $6_{1}$ and $6_{7}$ correspond to the $q=2$ Bose-Laughlin state and the first breathing mode, respectively. All quantities are dimensionless.}
\begin{ruledtabular}
\begin{tabular}{ccccccccccc}	
 && \multicolumn{2}{c}{$\mbox{g}_{2}=0.009151$} && \multicolumn{2}{c}{$\mbox{g}_{2}=0.09151$} && \multicolumn{2}{c}{$\mbox{g}_{2}=0.9151$} \\ 
\cline{3-4}\cline{6-7}\cline{9-10}\noalign{\smallskip} 
$i$ && $E(6_{i})$ & $S(6_{i})$ && $E(6_{i})$ & $S(6_{i})$ && $E(6_{i})$ & $S(6_{i})$ \\ \hline
{\blue 1} && {\blue 13.2426} & {\blue 1.5570} && {\blue 13.2426} & {\blue 1.5570} && {\blue 13.2426} & {\blue 1.5570} \\
 2 && 13.2434 & 1.6314 && 13.2503 & 1.6316 && 13.3149 & 1.6380 \\
 3 && 13.2436 & 1.2690 && 13.2522 & 1.2714 && 13.3349 & 1.3273 \\
 4 && 13.2437 & 1.5335 && 13.2528 & 1.5289 && 13.3409 & 1.5082 \\
 5 && 13.2440 & 1.6606 && 13.2566 & 1.6608 && 13.3792 & 1.6645 \\
 6 && 13.2455 & 1.5296 && 13.2714 & 1.5303 && 13.5158 & 1.5445 \\
{\blue 7} && {\blue 15.2426} & {\blue 2.3204} && {\blue 15.2426} & {\blue 2.3204} && {\blue 15.2426} & {\blue 2.3204} \\
 8 && 15.2429 & 2.2640 && 15.2451 & 2.2636 && 15.2684 & 2.2587 \\
 9 && 15.2434 & 2.1230 && 15.2498 & 2.1234 && 15.3152 & 2.1279 \\
10 && 15.2435 & 2.2826 && 15.2512 & 2.2784 && 15.3266 & 2.2492 \\
\end{tabular}
\end{ruledtabular}
\end{table}
\\
\indent
In order to measure the quantum correlation in variationally obtained states, in particular the breathing modes, we calculate the von Neumann (entanglement) entropy defined in terms of single-particle reduced density matrix $\hat{\rho}_{1}$ \cite{py01,lgc09,lf10} as 
\begin{equation}
S=-\mbox{Tr}\left(\hat{\rho}_{1} \ln \hat{\rho}_{1}\right)
\label{vne}
\end{equation}
in subspaces of total angular momentum $L$. 
In Table~\ref{tab:bql6}, we present the von Neumann entropy for $N=3$ bosons in $L=6$ subspace for three different values of interaction parameter. 
We observe that the value of the von Neumann entropy ($S$) for the first breathing mode $6_{7}$ is large compared to the Bose-Laughlin (ground) state $6_{1}$ in $L=6$ subspace. 
Surprisingly, we further note that the value of $S$ for both of these states ($6_{1}$ and $6_{7}$) remains unchanged as the interaction parameter is varied. 
It is, however, seen from Table~\ref{tab:bql6} that the eigenenergy and the corresponding von Neumann entropy of the  eigenstates lying between the Bose-Laughlin state $6_{1}$ and the first breathing mode $6_{7}$ change their values  as the interaction is varied, in contrast to the ground ({\it i.e.} the Bose-Laughlin) state and the first breathing mode\footnote{On the basis of our analysis for $N=3$ bosons in quantized total angular momentum $L=6$ subspace, the interaction independence of values of eigenenergy and von Neumann entropy for the ground state $6_{1}$ and the first breathing mode $6_{7}$ is suggestive of writing a variational ansatz for these states as 
\begin{equation}
\Psi_{\frac{q}{2}N(N-1)} \left(\left\{ z_{i}\right\} \right) \propto  \left(\prod_{i<j}\left(z_{i}-z_{j}\right)\right) \Psi_{\frac{q-1}{2}N(N-1)}^{CF}  \left(\left\{ z_{i}\right\} \right)
\label{cf}
\end{equation}
where $q=2,4, \cdots$, is an even positive integer (the normalization factor and the exponential factor symmetric in $\left\{z_{i}\right\}$ in the above expression have been omitted for the ease of writing). The Jastrow prefactor $\prod_{i<j}\left(z_{i}-z_{j}\right)$ with angular momentum $L_{JP}=\frac{1}{2}N\left(N-1\right)$ ensures interaction independence of the state $\Psi_{\frac{q}{2}N(N-1)} \left(\left\{ z_{i}\right\} \right)$. The Slater-determinant wavefunction $\Psi^{CF}_{\frac{q-1}{2}N(N-1)}$ for composite fermions with angular momentum $L_{CF}=\frac{q-1}{2}N\left(N-1\right)$ is to be constructed from single-particle basis states $\left\{u_{n,m}\right\}$, Eq.~(\ref{sps2}), from the lowest as well as higher Landau levels so that $\Psi_{\frac{q}{2}N(N-1)}\left(\left\{ z_{i}\right\} \right)$ has variational energy equal to the eigenenergy of the respective (the ground or the first breathing mode) states with total angular momentum $L=L_{JP}+L_{CF}=\frac{q}{2}N\left(N-1\right)$. For the unique choice of single-particle angular momentum states with $m=0,1,\dots , \left(N-1\right)$ in the lowest Landau levels, the Slater-determinant with $q=2$ becomes $\Psi_{\frac{1}{2}N(N-1)}^{CF}= \prod_{i<j}\left(z_{i}-z_{j}\right)$ and Eq.~(\ref{cf}) reduces to the Bose-Laughlin ground state wavefunction \cite{cw99} of Eq.~(\ref{bl}) for $q=2$. However, for the first breathing mode, a  prescription to uniquely construct $\Psi^{CF}_{\frac{q-1}{2}N(N-1)}$ with the above constraints on energy, is hard to find and may be determined variationally.}.
\\
\indent 
The internal structure (spatial correlation) of a many-body state can be analysed by calculating the conditional probability distribution (CPD) \cite{yl10,blo06,yl00} defined as
\begin{equation}
\mathcal{P} \left({\bf r},{\bf r}_{0}\right) = \frac{\langle \Psi\vert \sum_{i \neq j} \delta \left({\bf r}- {\bf r}_{i} \right)\delta \left({\bf r}_{0}-{\bf r}_{j}\right) \vert \Psi \rangle}{\left(N-1\right) \sum_{j} \langle \Psi \vert \delta \left({\bf r}_{0}-{\bf r}_{j}\right) \vert \Psi \rangle }
\label{cpd}
\end{equation}
where $\vert \Psi \rangle$ is the many-body eigenstate obtained through exact diagonalization and ${\bf r}_{0}=(x_{0},y_{0})$ is the reference point (usually chosen to be the position of high density for a few-body system like ours). 
The CPD can be interpreted as the probability of a particle being at position ${\bf r}$ under the condition that another one is located at ${\bf r}_{0}$. 
\begin{figure*}[!htb]
\centering
\subfigure[$~6_{1}$ BL state]{\includegraphics[width=0.19\linewidth]{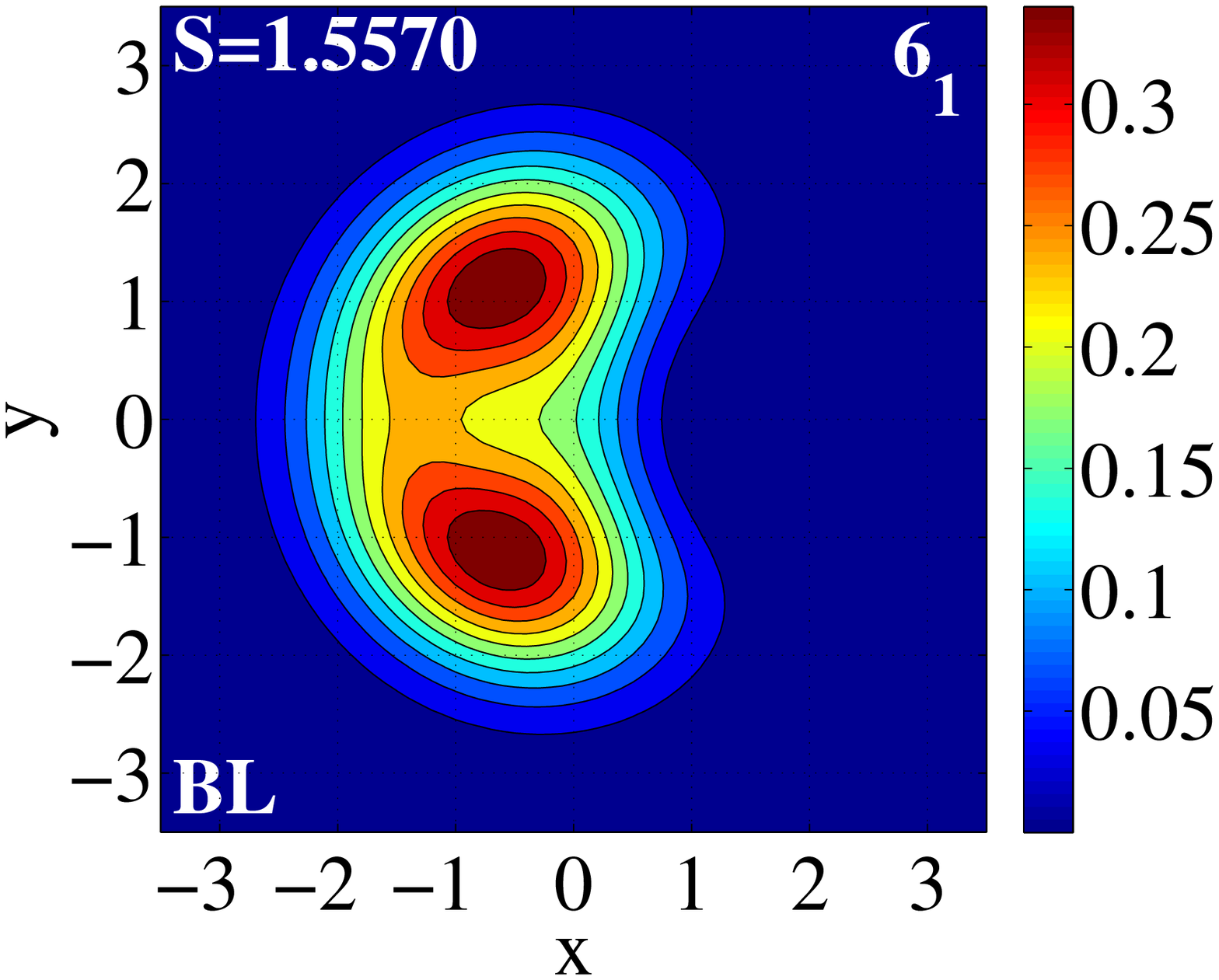}\label{fig:cpdn3l6.sub1}}
\subfigure[$~6_{2}$]{\includegraphics[width=0.19\linewidth]{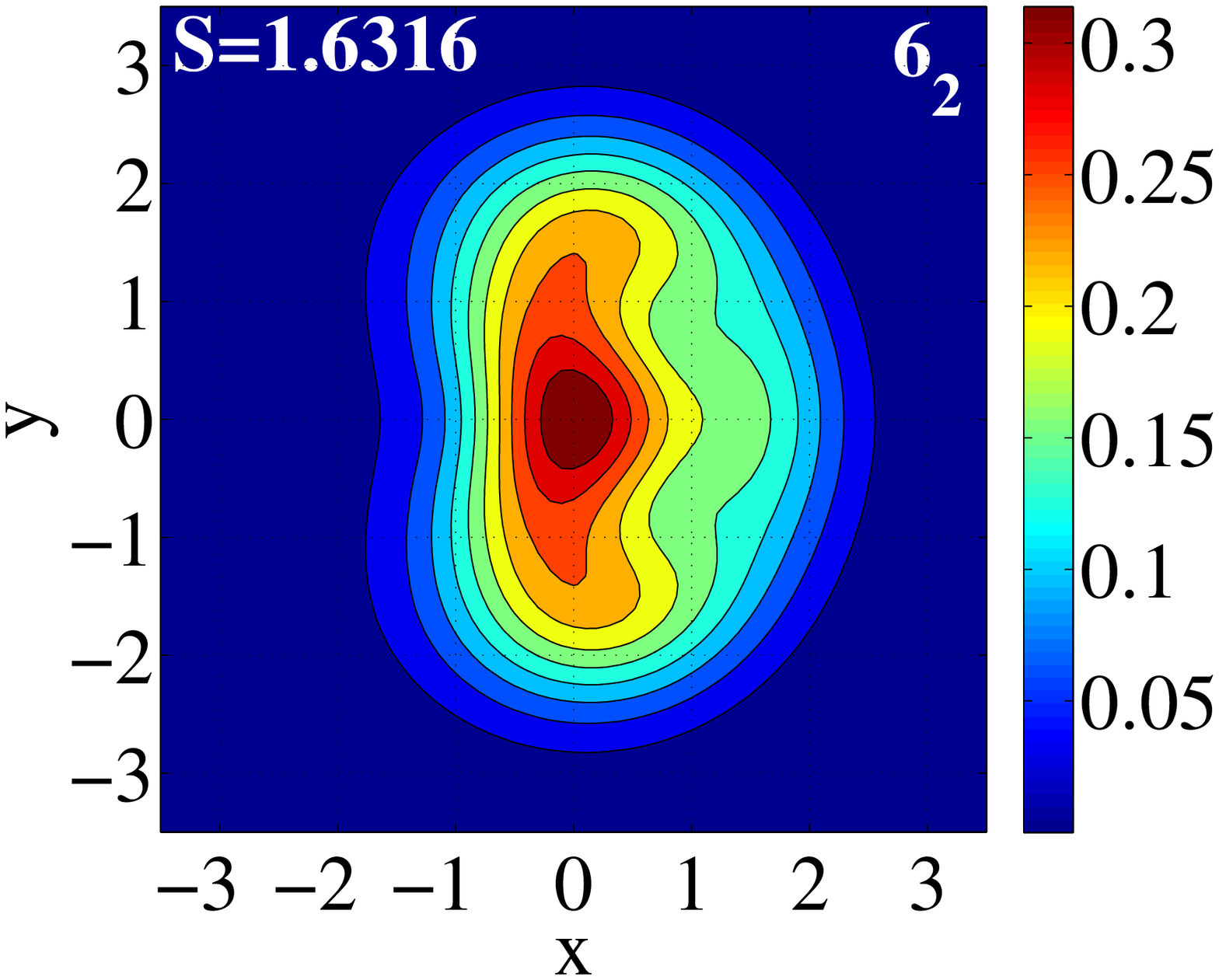}\label{fig:cpdn3l6.sub2}}
\subfigure[$~6_{3}$]{\includegraphics[width=0.19\linewidth]{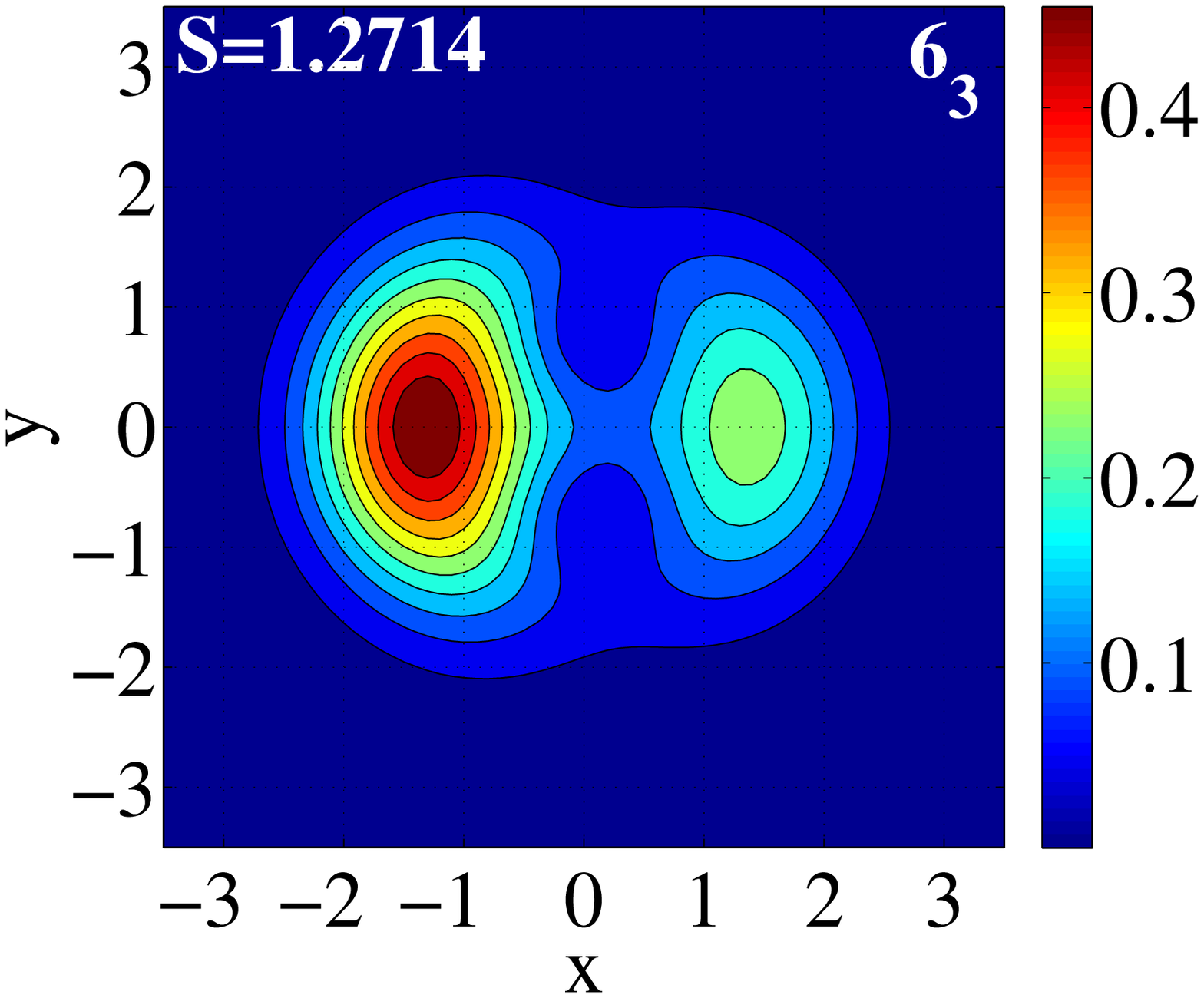}\label{fig:cpdn3l6.sub3}}
\subfigure[$~6_{4}$]{\includegraphics[width=0.19\linewidth]{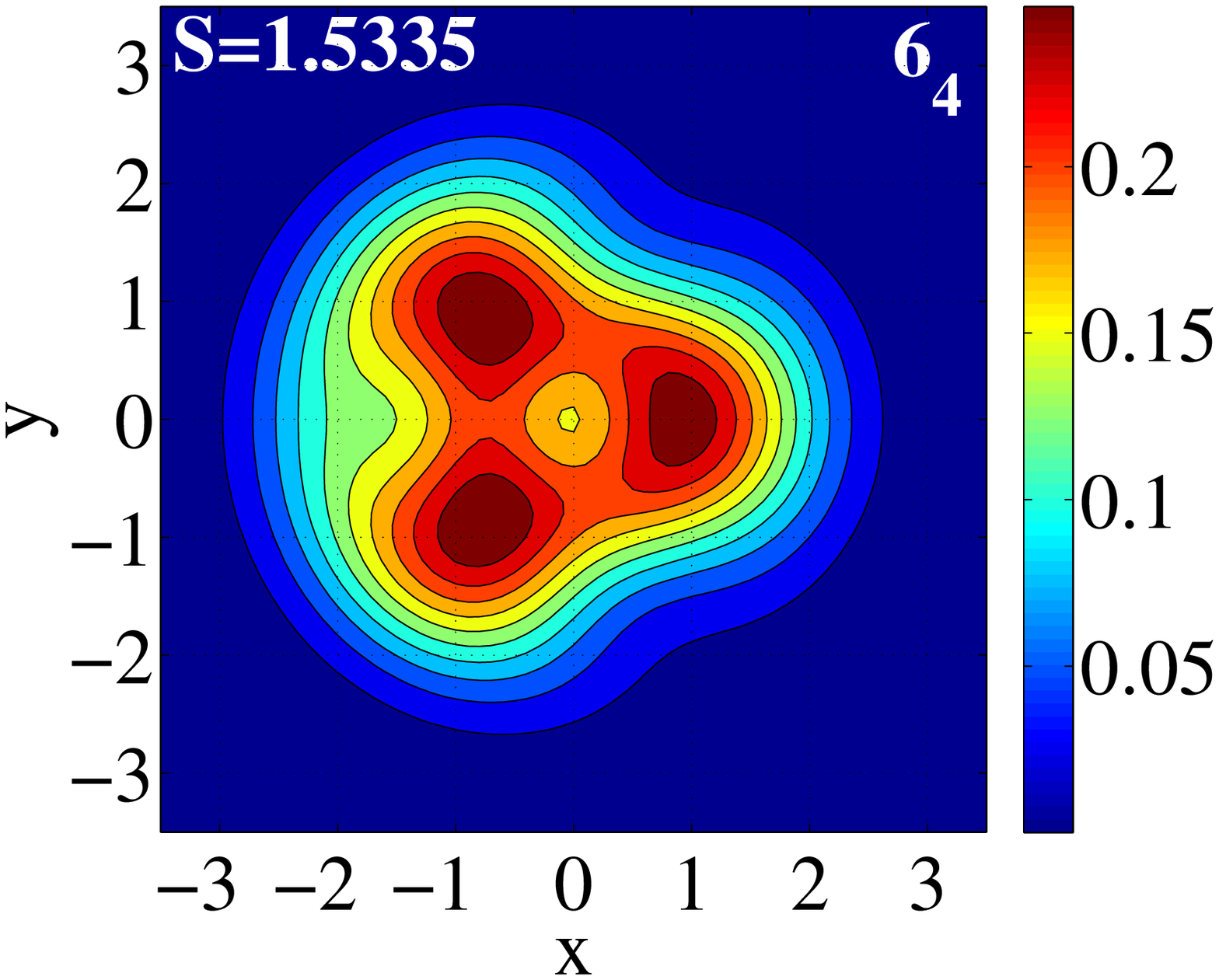}\label{fig:cpdn3l6.sub4}}
\subfigure[$~6_{5}$]{\includegraphics[width=0.19\linewidth]{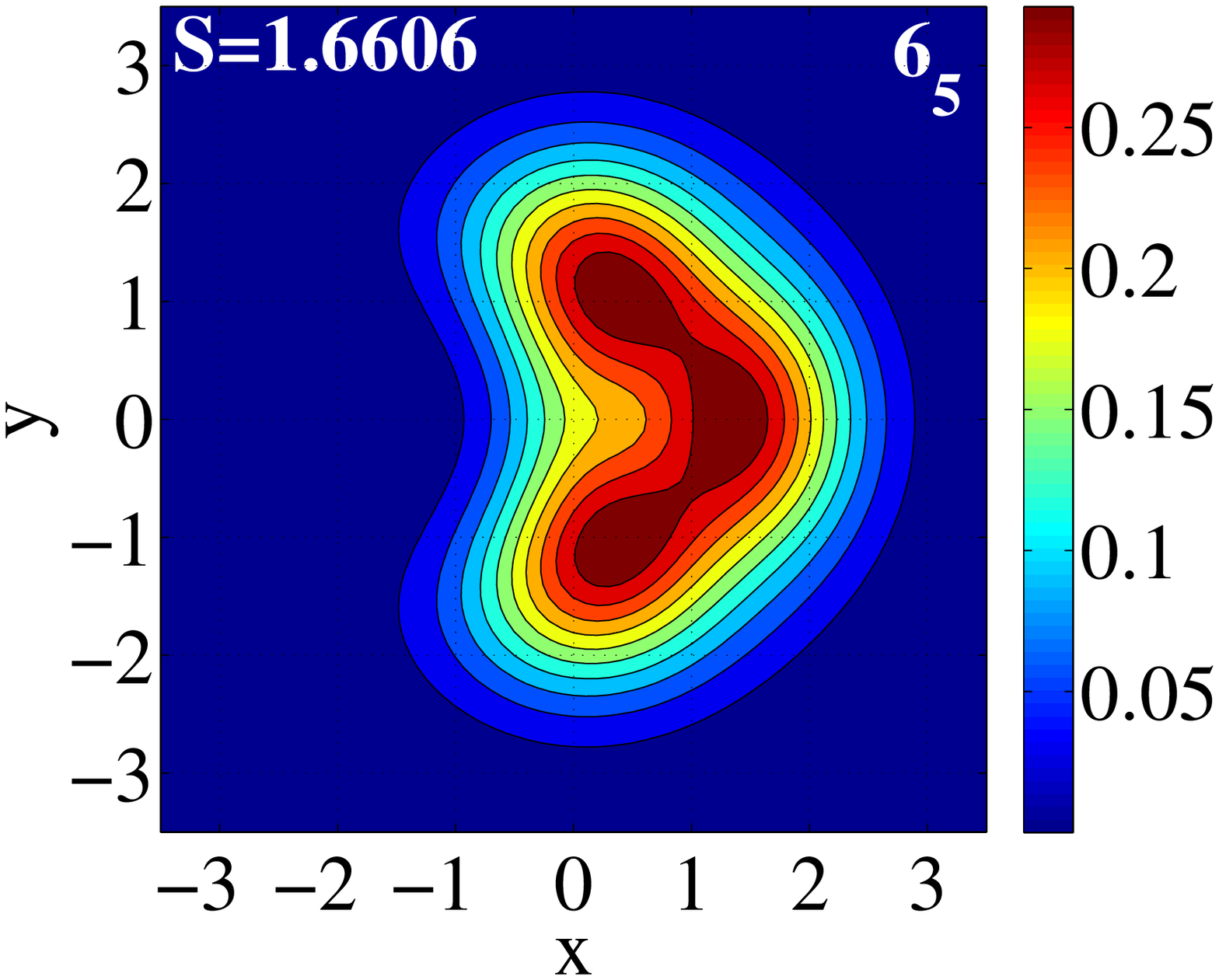}\label{fig:cpdn3l6.sub5}}
\subfigure[$~6_{6}$]{\includegraphics[width=0.19\linewidth]{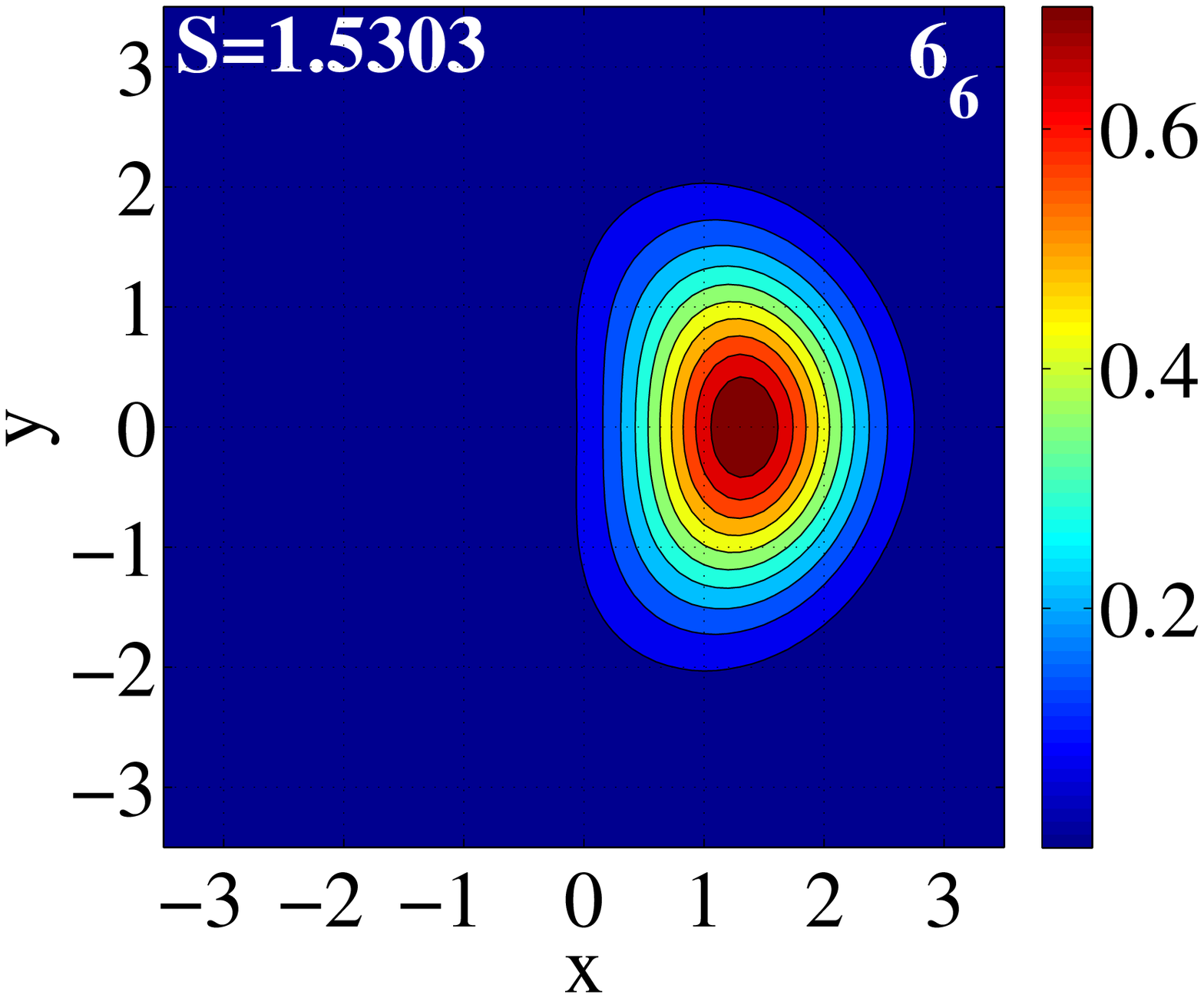}\label{fig:cpdn3l6.sub6}}
\subfigure[$~6_{7}$ BM]{\includegraphics[width=0.19\linewidth]{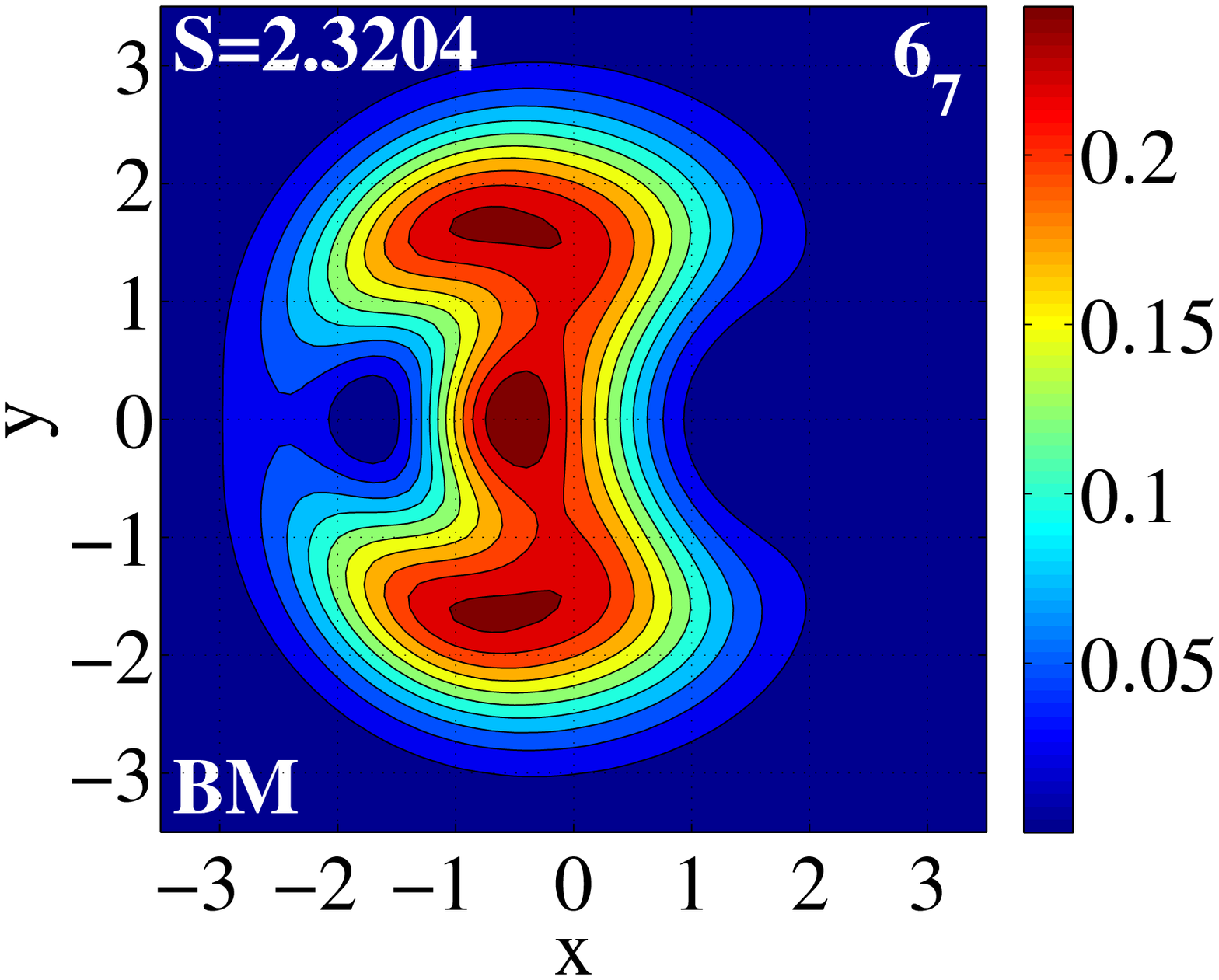}\label{fig:cpdn3l6.sub7}}
\subfigure[$~6_{8}$]{\includegraphics[width=0.19\linewidth]{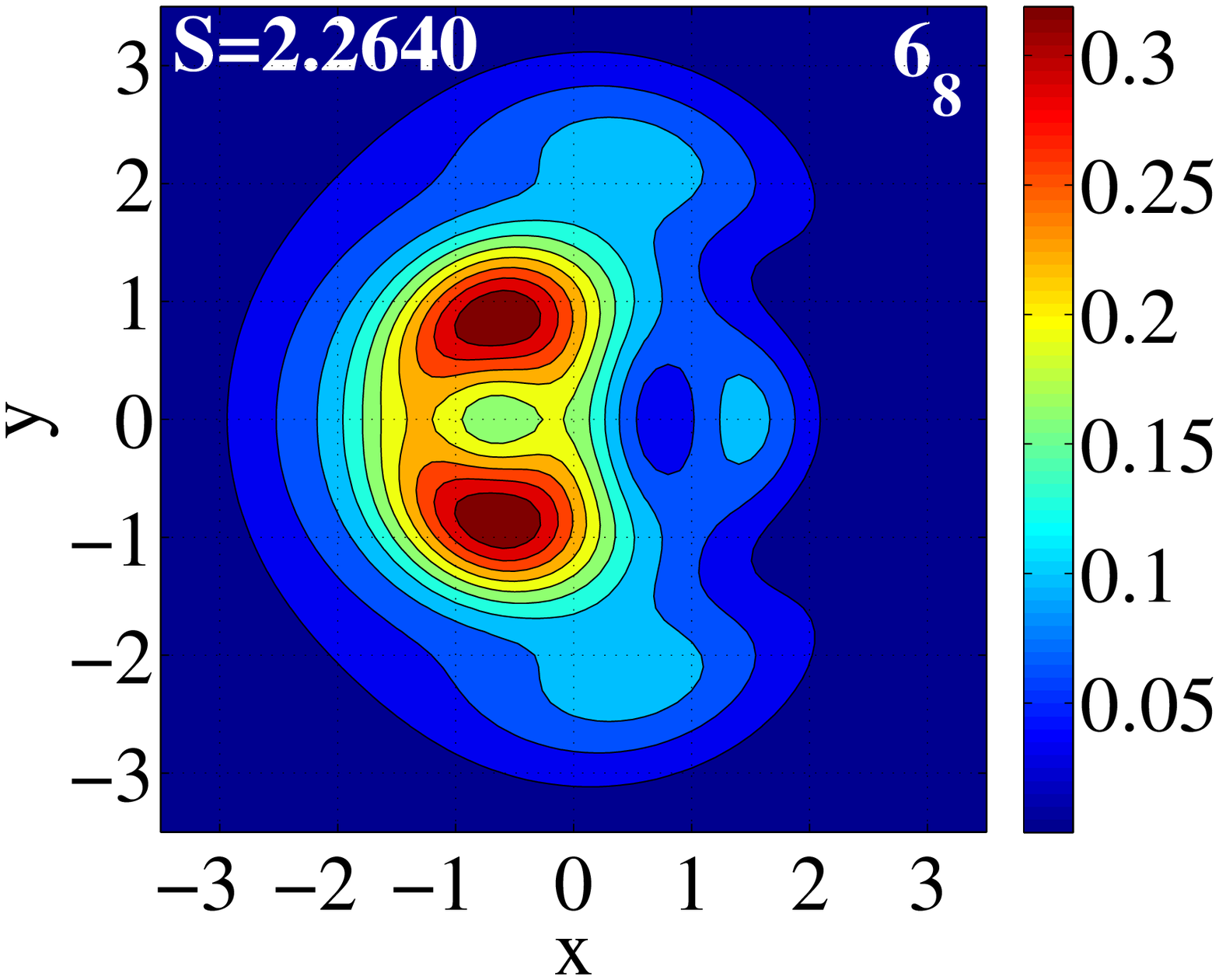}\label{fig:cpdn3l6.sub8}}
\subfigure[$~6_{9}$]{\includegraphics[width=0.19\linewidth]{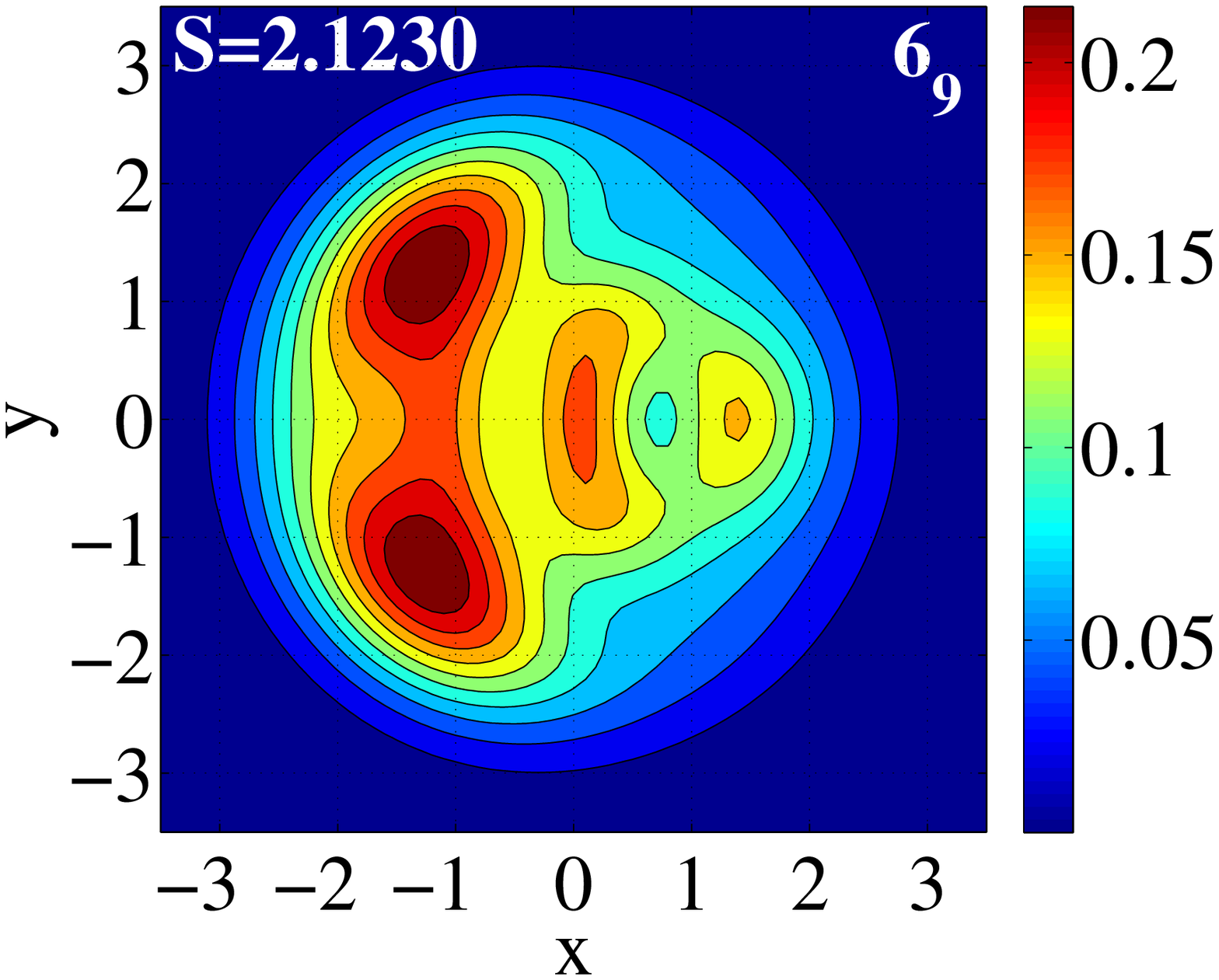}\label{fig:cpdn3l6.sub9}}
\subfigure[$~6_{10}$]{\includegraphics[width=0.19\linewidth]{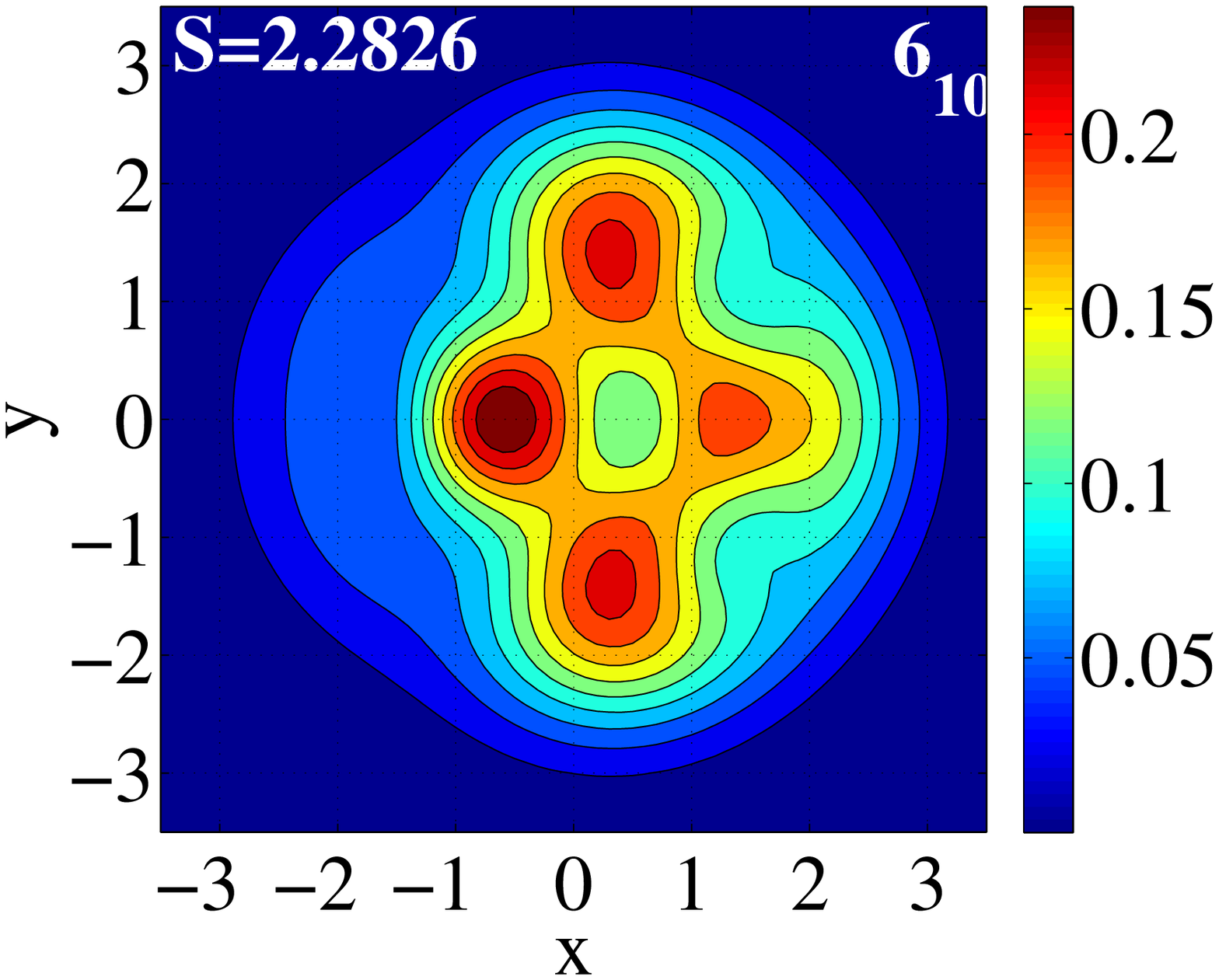}\label{fig:cpdn3l6.sub10}}
\caption{\label{fig:cpdn3l6}(Color online) Contour plots for conditional probability distribution (CPD) of low-lying eigenstates in angular momentum subspace $L=6$ for $N=3$ bosons with $\mbox{g}_{2}=0.09151$ and $\sigma_{\perp}=0.1$ in Eq.~(\ref{gip}). The plots (isosurface density profiles viewed along $z$-axis) show the probability distribution of finding a particle at position $(x,y)$ when another particle has been fixed at a position of relatively high density chosen to be ${\bf r}_{0}=(x_{0},y_{0})=\left(1.732,0\right)$, here. Brown-red regions have the highest probability density falling off to blue regions of low probability density.}
\end{figure*}
\\
\indent
In Fig.~\ref{fig:cpdn3l6}, we present CPD plots of low-lying eigenstates for $N=3$ in angular momentum subspace $L=6$ for the interaction parameter $\mbox{g}_{2}=0.09151$ and $\sigma_{\perp}=0.1$ in Eq.~(\ref{gip}) with  reference point chosen at ${\bf r}_{0}=\left(\sqrt{N},0\right)$ in units of $a_{\perp}$ in the $x$-$y$ plane.  
The ground state $6_{1}$ with two equal peaks in CPD, symmetrically placed with respect to ${\bf r}_{0}$ in Fig.~\ref{fig:cpdn3l6.sub1} exhibits a strong anticorrelation (exclusion) structure implying that the probability of finding two or more particles at the same position is vanishingly small. 
This underlines the composite fermion structure of the Bose-Laughlin state (\ref{bl}). 
In Fig.~\ref{fig:cpdn3l6.sub2} for the state $6_{2}$, the anticorrelation structure vanishes as two of the peaks of state $6_{1}$, merge around the trap center to form a higher peak. 
The merged peak around the trap centre in CPD of state $6_{2}$ in Fig.~\ref{fig:cpdn3l6.sub2}, re-distributes itself into two distinct unequal peaks in the state $6_{3}$ in Fig.~\ref{fig:cpdn3l6.sub3} where the relatively higher peak shifts away from the trap center and a smaller peak appears at the opposite end. 
In the state $6_{4}$ shown in Fig.~\ref{fig:cpdn3l6.sub4}, the higher peak of state $6_{3}$ splits into two equal peaks and the smaller peak becomes equally prominent with the three strongly correlated peaks forming an equilateral triangle. 
We observe that Figs.~\ref{fig:cpdn3l6.sub2} through \ref{fig:cpdn3l6.sub5}, the probability of finding one or more particle at the reference position ${\bf r}_{0}$ increases progressively implying increasing tendency towards bosonic correlation. 
In fact from the CPD plot in Fig.~\ref{fig:cpdn3l6.sub6} for the state $6_{6}$ with only one peak of about $0.6$ at ${\bf r}_{0}$, it appears as if all the bosons occupy the same position resulting in a contracted state exhibiting peaked bosonic correlation. 
Interestingly, the state $6_{6}$ is precursor to the expanded state $6_{7}$ identified as the lowest eigenstate of the first breathing band and has internal structure shown in Fig.~\ref{fig:cpdn3l6.sub7} with two equal peaks at the opposite ends and one right at the trap center. 
We speculate that this first breathing mode $6_{7}$ possibly has structure similar to the state in Eq.~(\ref{cf}). 
CPD plots in Figs.~\ref{fig:cpdn3l6.sub8}-\ref{fig:cpdn3l6.sub10} which belong to the first breathing band \cite{bhh02} can interpreted on the same line.   
\begin{figure}[!htb]
\centering
\subfigure[$~\mbox{g}_{2}=0.009151$]{{\includegraphics[width=0.46\linewidth]{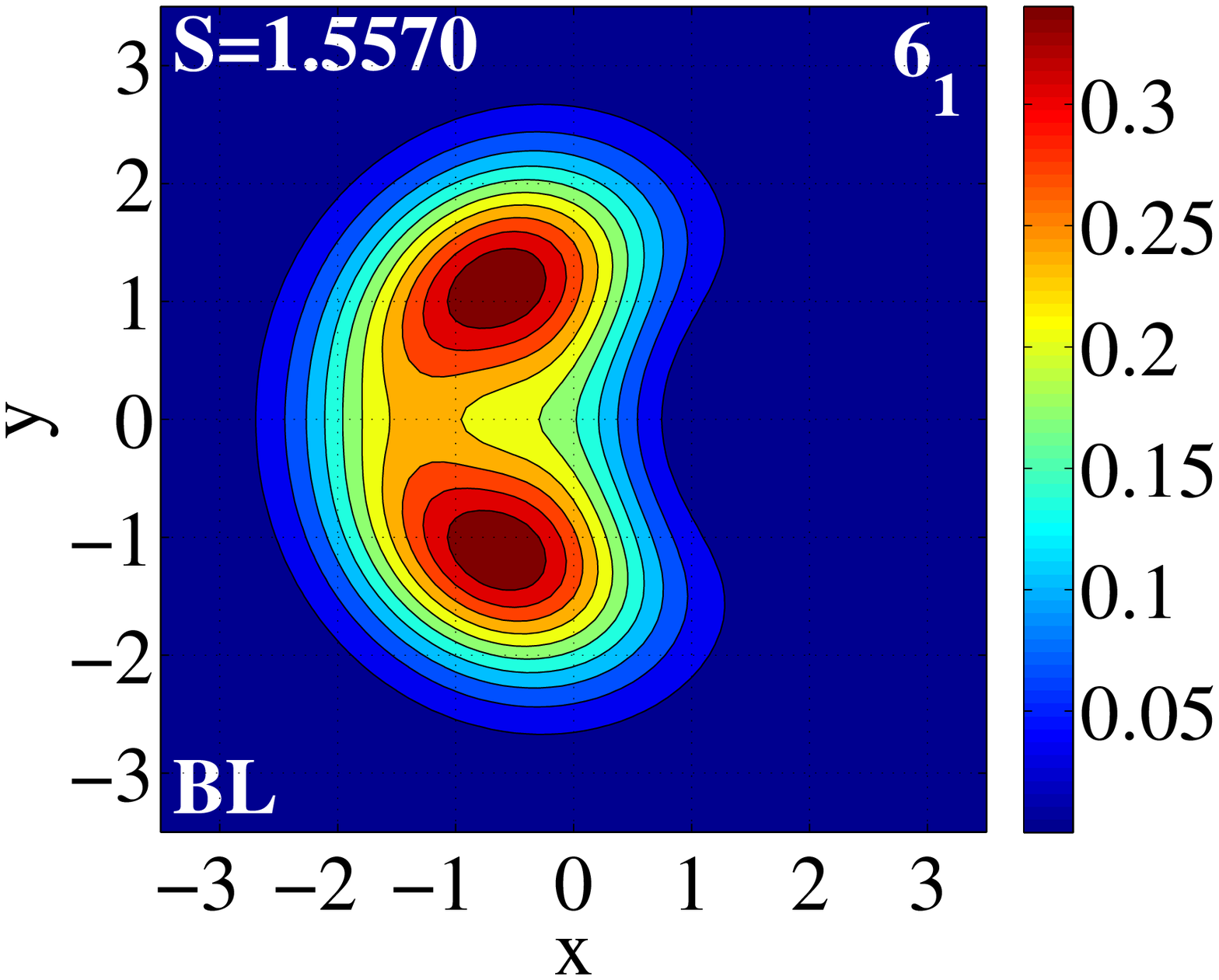}} {\includegraphics[width=0.46\linewidth]{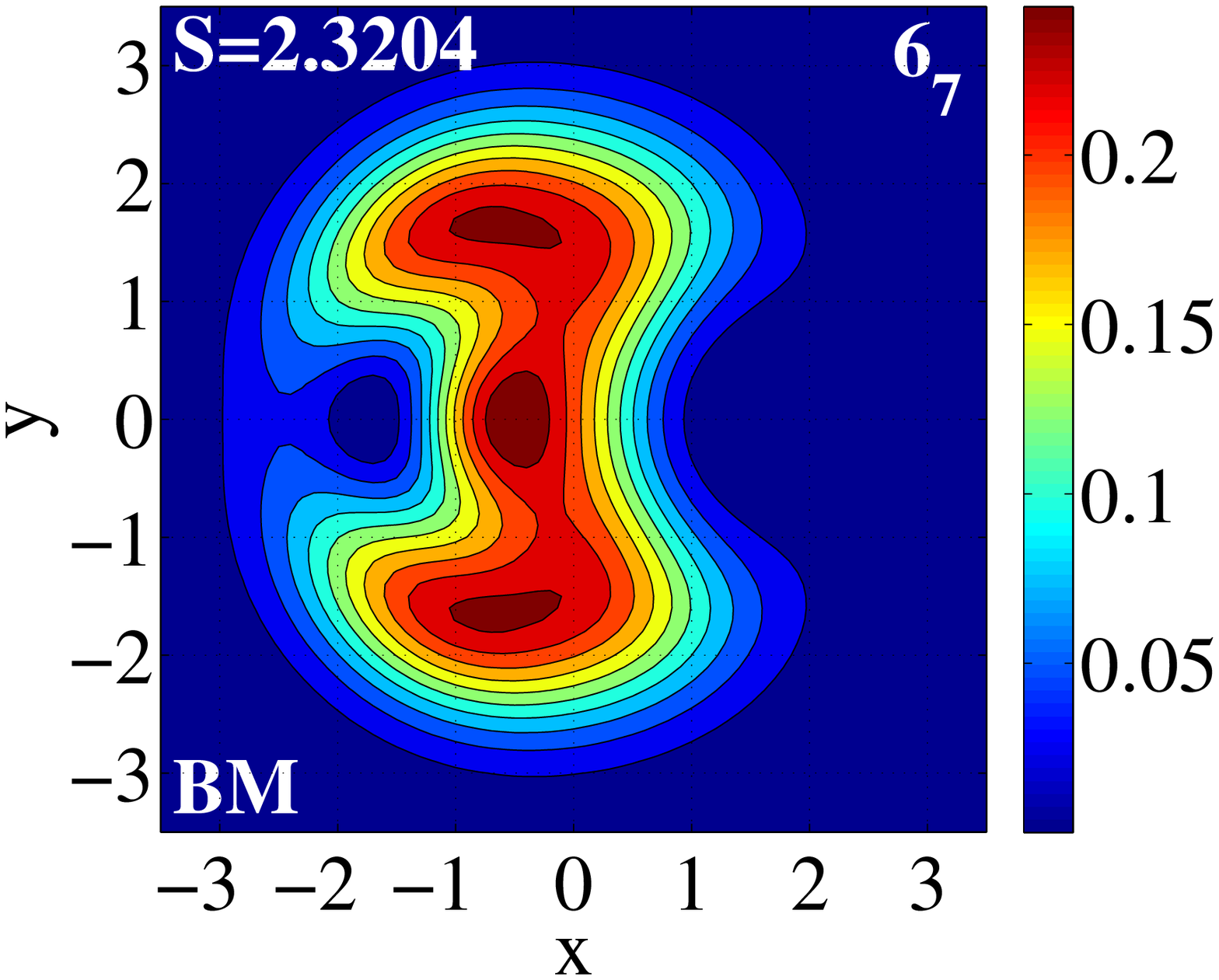}}\label{fig:cpdbl6.sub1}}
\subfigure[$~\mbox{g}_{2}=0.9151$]{{\includegraphics[width=0.46\linewidth]{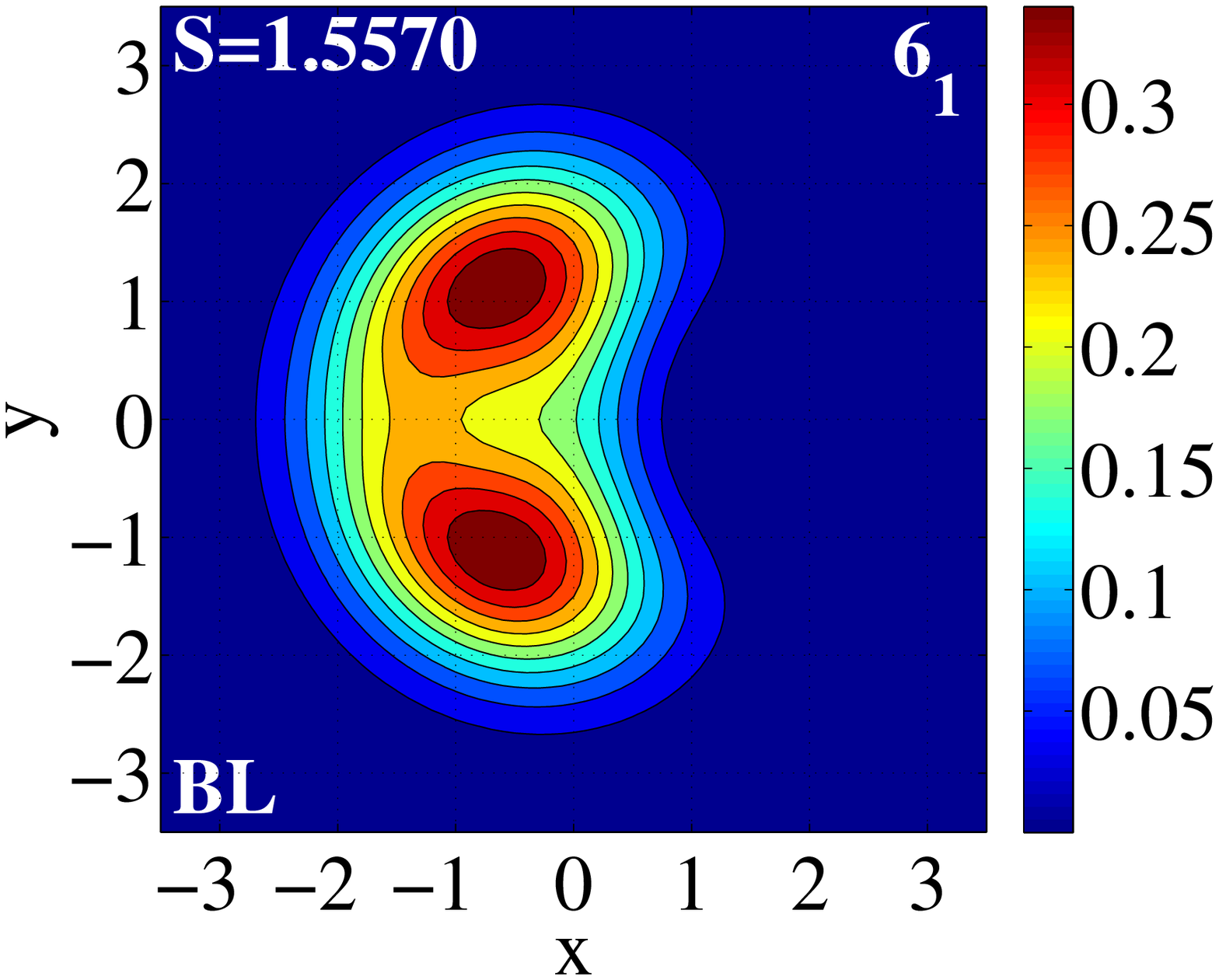}} {\includegraphics[width=0.46\linewidth]{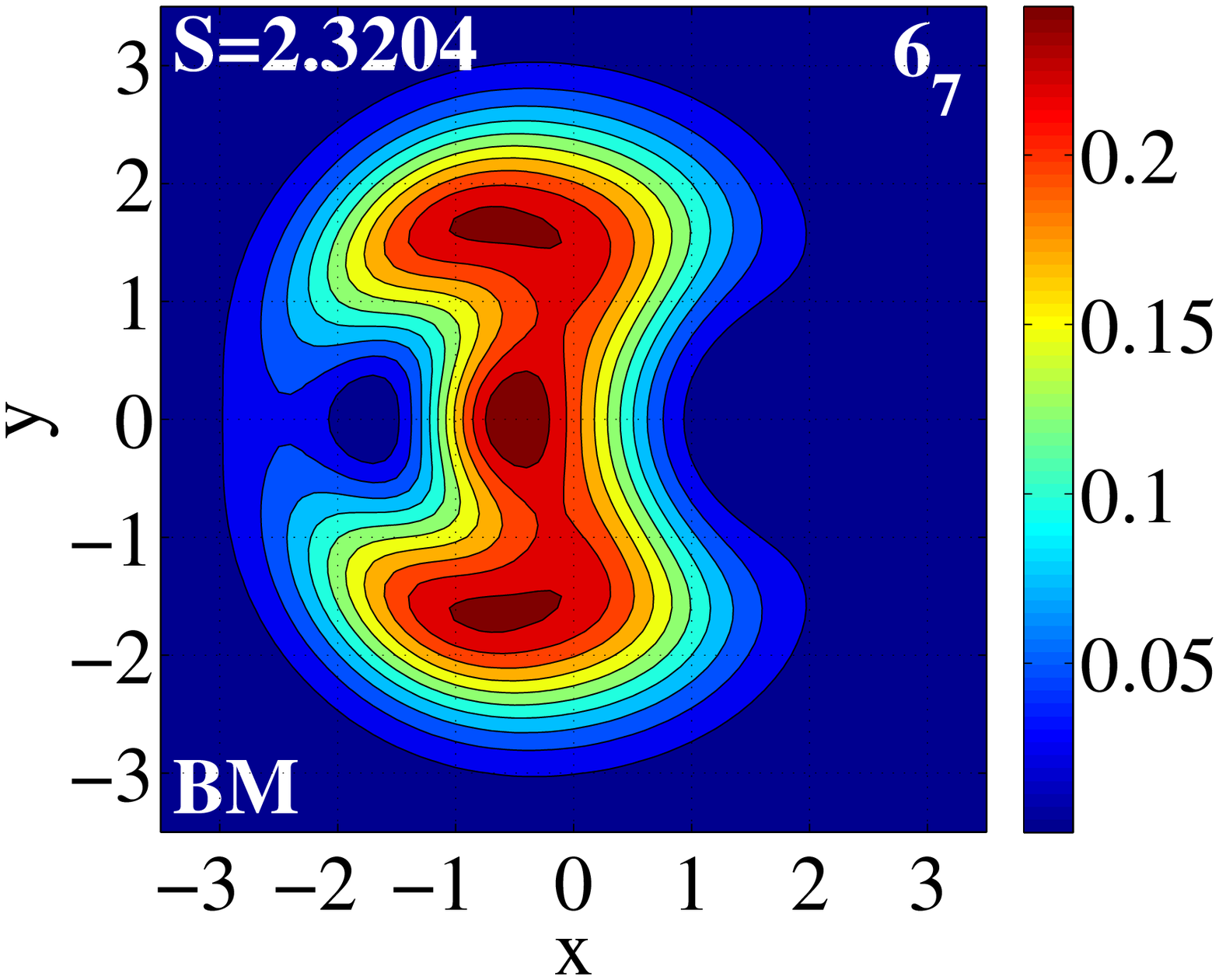}}\label{fig:cpdbl6.sub2}}
\caption{\label{fig:cpdbl6}(Color online) For $N=3$ bosons in angular momentum $L=6$ subspace, the conditional probability distribution (CPD) plots of the $q=2$ Bose-Laughlin (ground) state $6_{1}$ and the first breathing mode $6_{7}$ for interaction parameter (a) $\mbox{g}_{2}=0.009151$, (b) $\mbox{g}_{2}=0.9151$ with Gaussian width $\sigma_{\perp}=0.1$ in Eq.~(\ref{gip}). In all CPD plots, brown-red regions have the highest probability (of finding a particle) falling off to blue regions of low probability.}
\end{figure}
\\
\indent
In order to examine the effect of repulsive interaction on the first breathing mode, we present CPD plots in Fig.~\ref{fig:cpdbl6} for angular momentum subspace $L=6$ with two values of interaction parameter $\mbox{g}_{2}=0.009151$ and $0.9151$. 
First, like the eigenenergy and the von Neumann entropy, the internal structure of the ground state $6_{1}$ too is independent of interaction as seen in Fig.~\ref{fig:cpdbl6.sub1} and Fig.~\ref{fig:cpdbl6.sub2}.
This is consistent with the very form of the Bose-Laughlin state~(\ref{bl}) which does not allow two particles being in the same position. 
We further observe that the internal structure of the first breathing mode $6_{7}$ remains unchanged for the two values of interaction parameter in Fig.~\ref{fig:cpdbl6.sub1} and Fig.~\ref{fig:cpdbl6.sub2}.
This is corroborated by invariant values of the von Neumann entropy $(S=2.3204)$, given at top left in each plot. 
We thus, observe that the $q=2$ Bose-Laughlin state and the corresponding first breathing mode in $L=6$ subspace are similar in having the eigenenergy, the von Neumann entropy and the internal structure independent of interaction. 
\begin{table}[!htb]
\caption{\label{tab:bql12}For $N=3$ rapidly rotating bosons in total angular momentum subspace $L=12$, values of eigenenergy ($E$) and von Neumann entropy ($S$) of the ground state and low-lying excited states including the first breathing mode with interaction parameters $\mbox{g}_{2}=0.009151$, $0.09151$, $0.9151$ and range $\sigma_{\perp}=0.1$ of the Gaussian potential~(\ref{gip}). The states ${12}_{1}$ and ${12}_{9}$ correspond to the $q=4$ Bose-Laughlin state and the first breathing mode, respectively. All quantities are dimensionless.}
\begin{ruledtabular}
\begin{tabular}{cccccccccc}
&& \multicolumn{2}{c}{$\mbox{g}_{2}=0.009151$} && \multicolumn{2}{c}{$\mbox{g}_{2}=0.09151$} && \multicolumn{2}{c}{$\mbox{g}_{2}=0.9151$} \\ 
\cline{3-4}\cline{6-7}\cline{9-10}\noalign{\smallskip} 
$i$ && $E({12}_{i})$ & $S({12}_{i})$ && $E({12}_{i})$ & $S({12}_{i})$ && $E({12}_{i})$ & $S({12}_{i})$ \\ \hline
{\blue 1} && {\blue 19.2426} & {\blue 1.7014} && {\blue 19.2426} & {\blue 1.7014} && {\blue 19.2426} & {\blue 1.7014} \\
2 && 19.2426 & 1.7962 && 19.2426 & 1.7961 && 19.2426 & 1.7962 \\
3 && 19.2434 & 1.8970 && 19.2503 & 1.8971 && 19.3192 & 1.8977 \\
4 && 19.2435 & 1.8965 && 19.2512 & 1.8966 && 19.3260 & 1.8961 \\
5 && 19.2435 & 1.6963 && 19.2518 & 1.6967 && 19.3326 & 1.6944 \\
6 && 19.2436 & 1.5044 && 19.2522 & 1.5067 && 19.3364 & 1.5159 \\
7 && 19.2437 & 1.7961 && 19.2542 & 1.7959 && 19.3570 & 1.7953 \\
8 && 19.2454 & 1.7515 && 19.2704 & 1.7520 && 19.5110 & 1.7593 \\
{\blue 9} && {\blue 21.2426} & {\blue 2.2549} && {\blue 21.2426} & {\blue 2.2549} && {\blue 21.2426} & {\blue 2.2549} \\
10 && 21.2426 & 2.2923 && 21.2429 & 2.2922 && 21.2459 & 2.2917 \\
\end{tabular}
\end{ruledtabular}
\end{table}
\begin{figure}[!htb]
\centering
\subfigure[$~\mbox{g}_{2}=0.009151$]{{\includegraphics[width=0.46\linewidth]{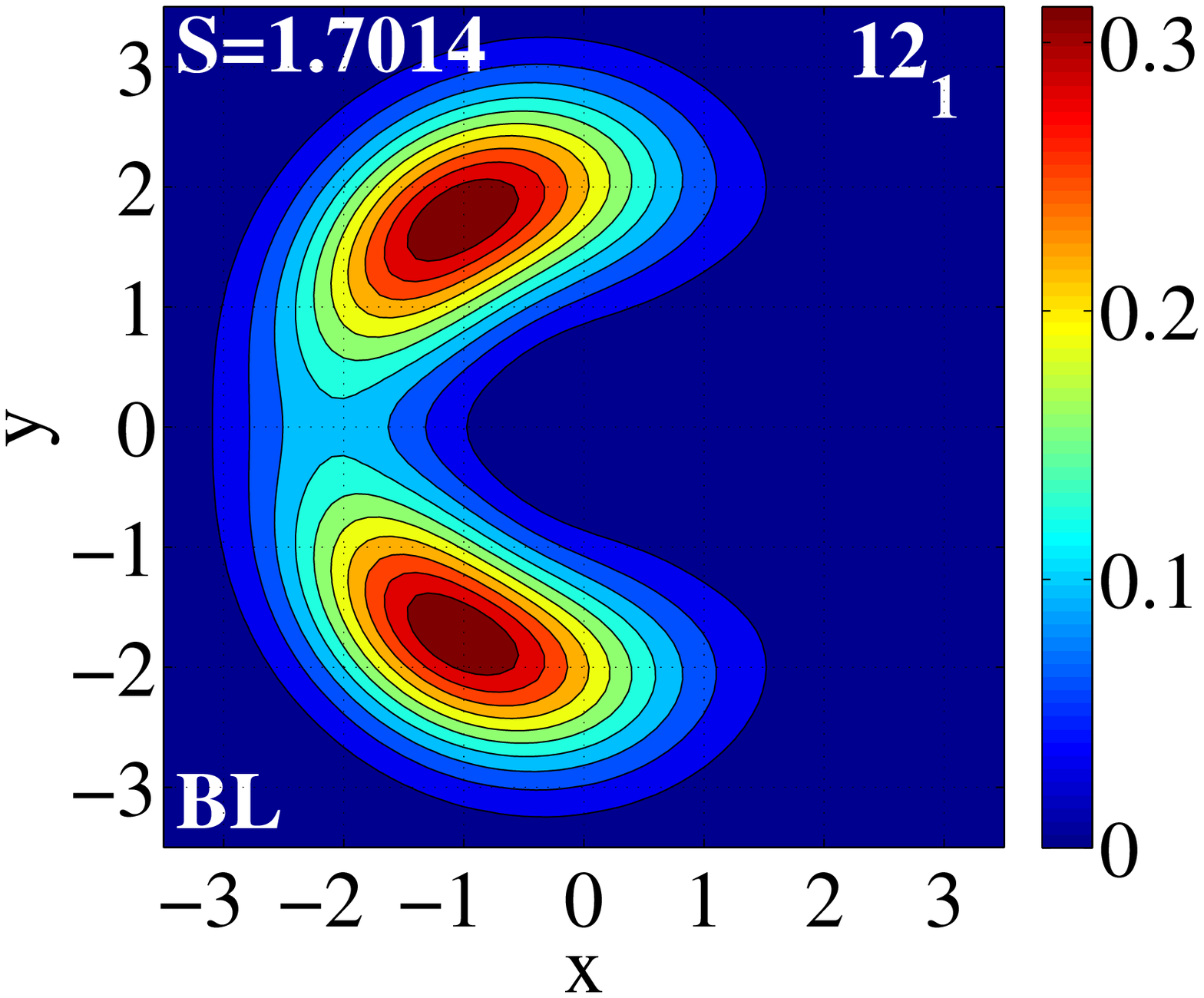}} {\includegraphics[width=0.46\linewidth]{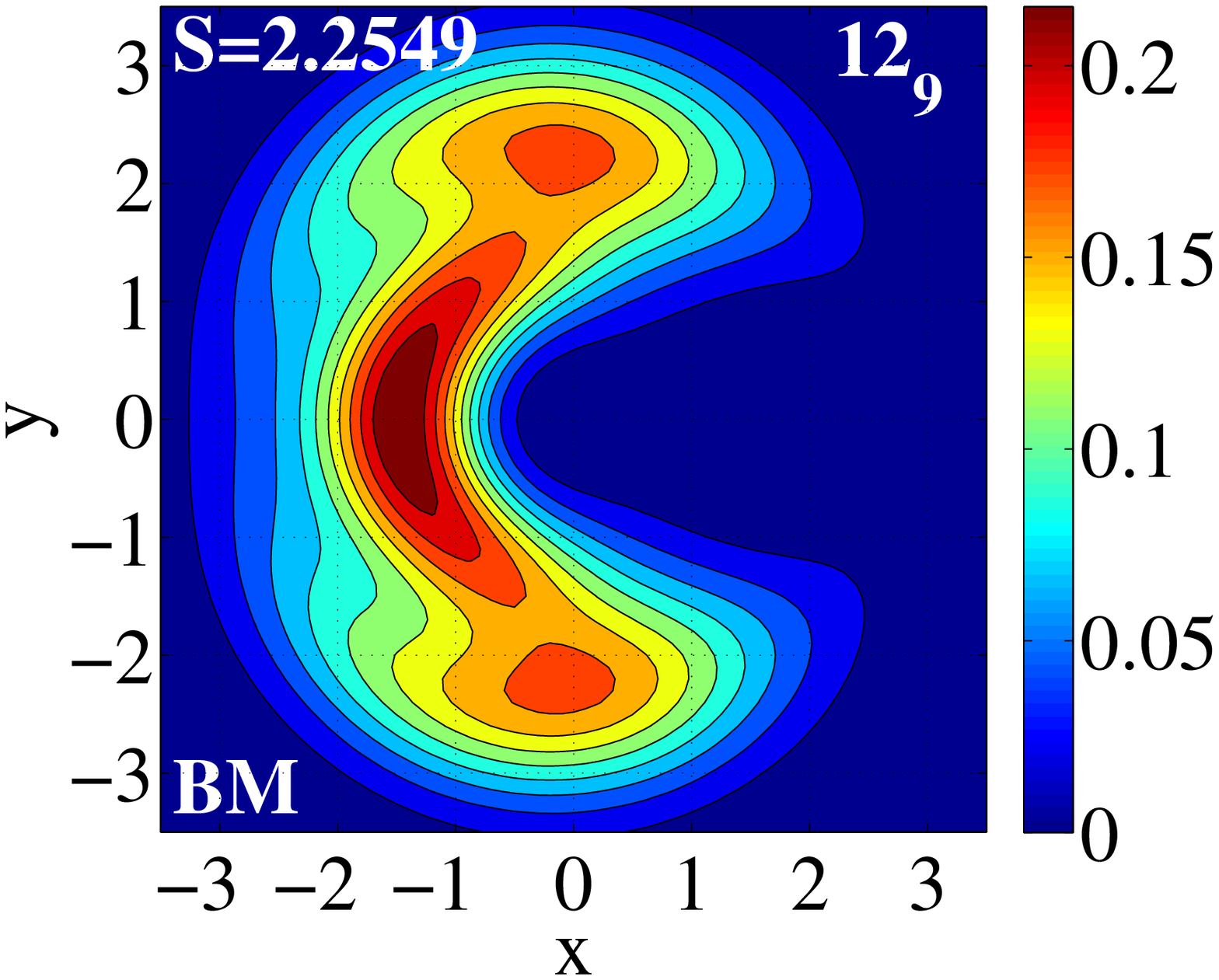}}\label{fig:cpdbl12.sub1}}
\subfigure[$~\mbox{g}_{2}=0.9151$]{{\includegraphics[width=0.46\linewidth]{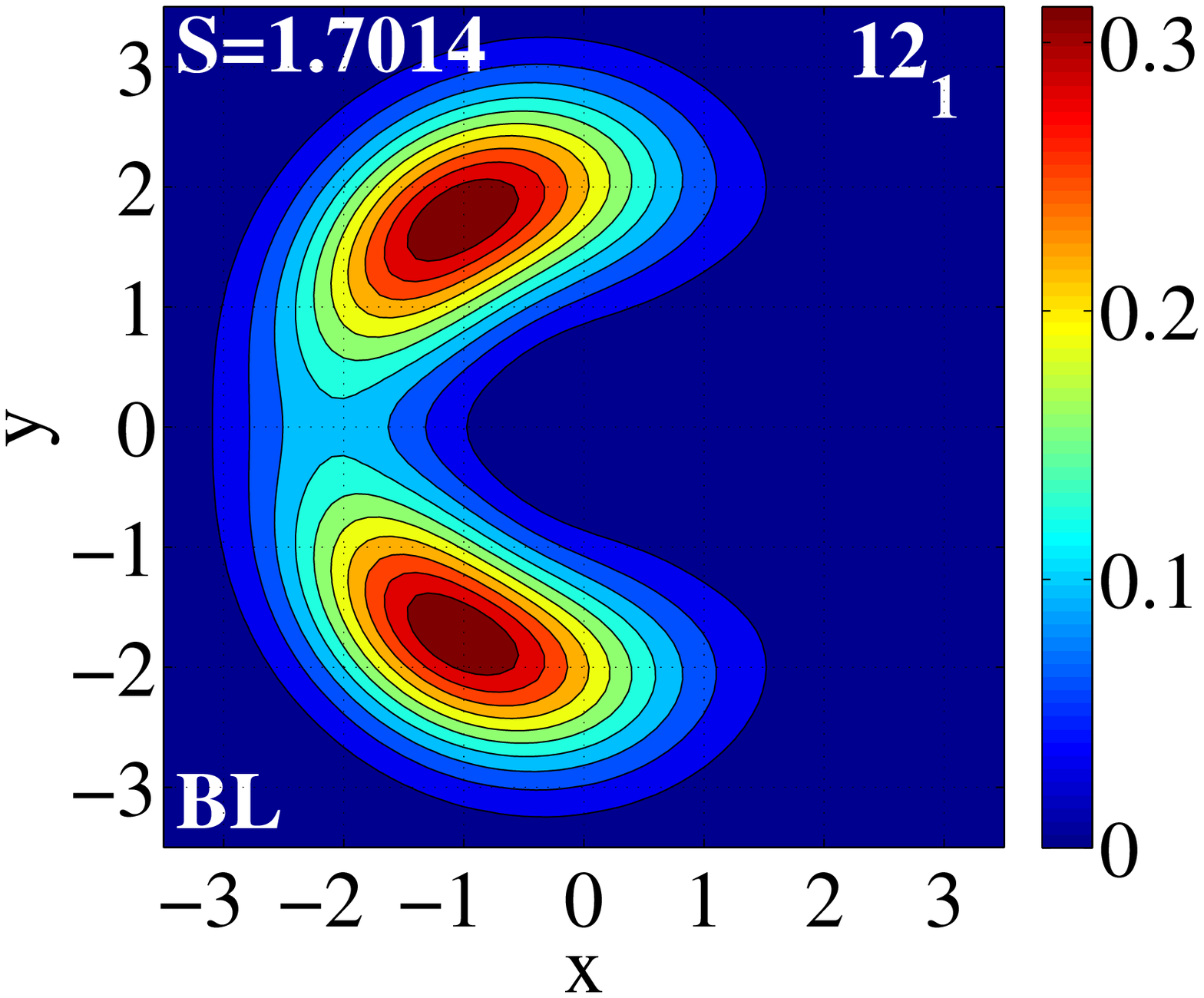}} {\includegraphics[width=0.46\linewidth]{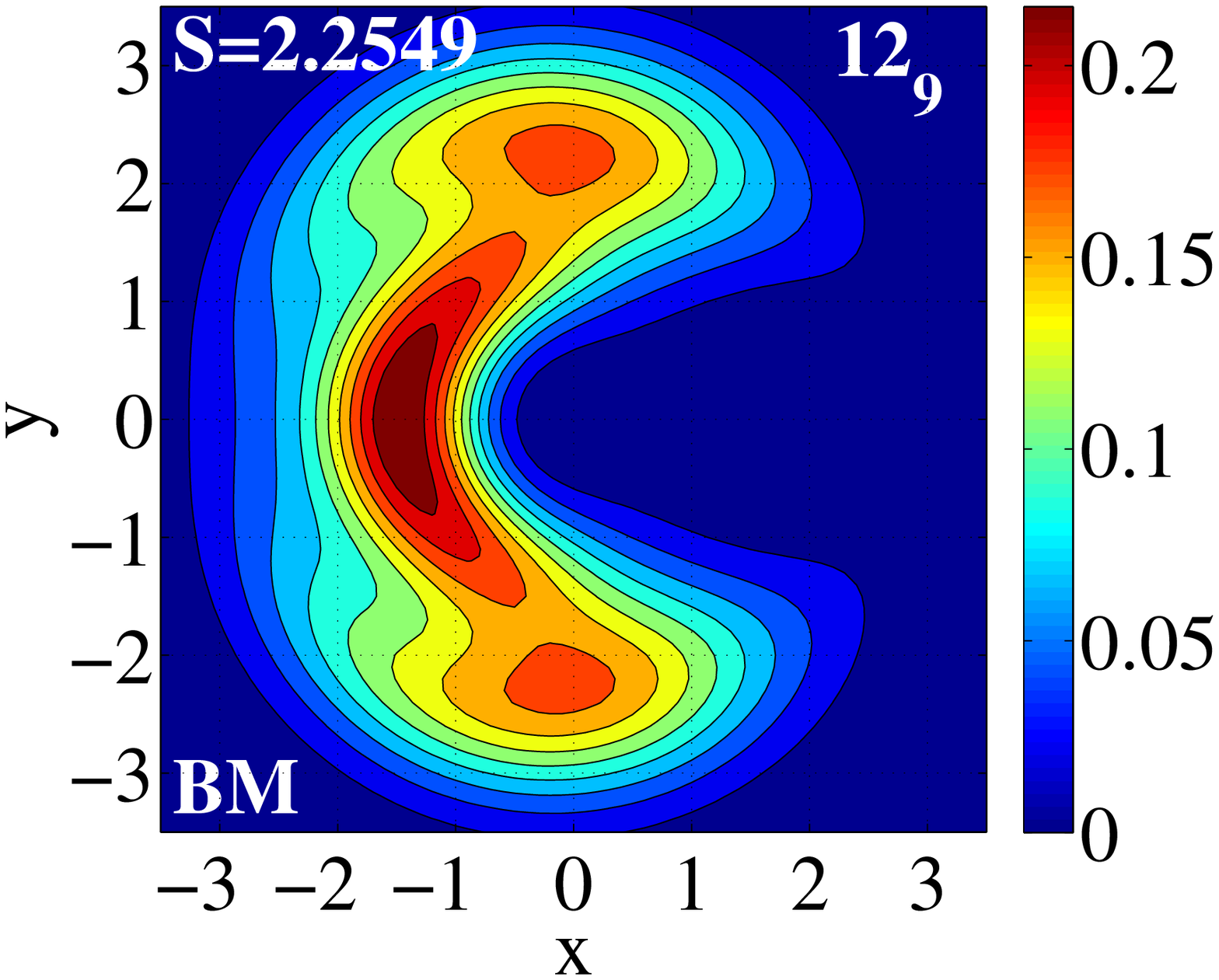}}\label{fig:cpdbl12.sub2}}
\caption{\label{fig:cpdbl12}(Color online) For $N=3$ bosons with total angular momentum $L=12$, the conditional probability distribution (CPD) of the $q=4$ Bose-Laughlin state ${12}_{1}$ and the first breathing mode ${12}_{9}$ for the interaction parameter (a) $\mbox{g}_{2}=0.009151$, (b) $\mbox{g}_{2}=0.9151$ with Gaussian width $\sigma_{\perp}=0.1$ in Eq.~(\ref{gip}). In all CPD plots, brown-red regions have highest probability (of finding a particle) falling off to blue regions of low probability. The internal structure as well as the von Neumann entropy (marked on the top left corner in each plot) of the Bose-Laughlin state ${12}_{1}$ and the first breathing mode ${12}_{9}$ are exactly the same even as the interaction parameter is varied over two orders of magnitude.}
\end{figure}
\\
\indent
The interaction independence of the eigenenergy, the von Neumann entropy and the internal structure is also seen in the $q=4$ Bose-Laughlin state and the corresponding first breathing mode in angular momentum subspace $L=12$. 
The diagonalization is performed for three different values of interaction parameter $\mbox{g}_{2}=0.009151$, $0.09151$ and $0.9151$, and the results are presented in Table~\ref{tab:bql12}.  
It is seen from the table that $E({12}_{9})-E({12}_{1})=2$, for the three values of interaction parameter considered. The states ${12}_{1}$ and ${12}_{9}$ are the $q=4$ Bose-Laughlin state and the first breathing mode respectively in $L=12$ subspace.
We further observe that the eigenenergy, the von Neumann entropy as well as internal structure (see Fig.~\ref{fig:cpdbl12}) of ${12}_{1}$ and ${12}_{9}$ states remain unchanged as the interaction parameter is varied over three orders of magnitude. 
However, values of eigenenergy and von Neumann entropy of the eigenstates lying between the Bose-Laughlin state $12_{1}$ and the first breathing mode $12_{9}$, vary with interaction as is seen from the Table~\ref{tab:bql12}.
\begin{figure*}[!]
\centering
\subfigure[$~{12}_{1}$ BL state]{\includegraphics[width=0.19\linewidth]{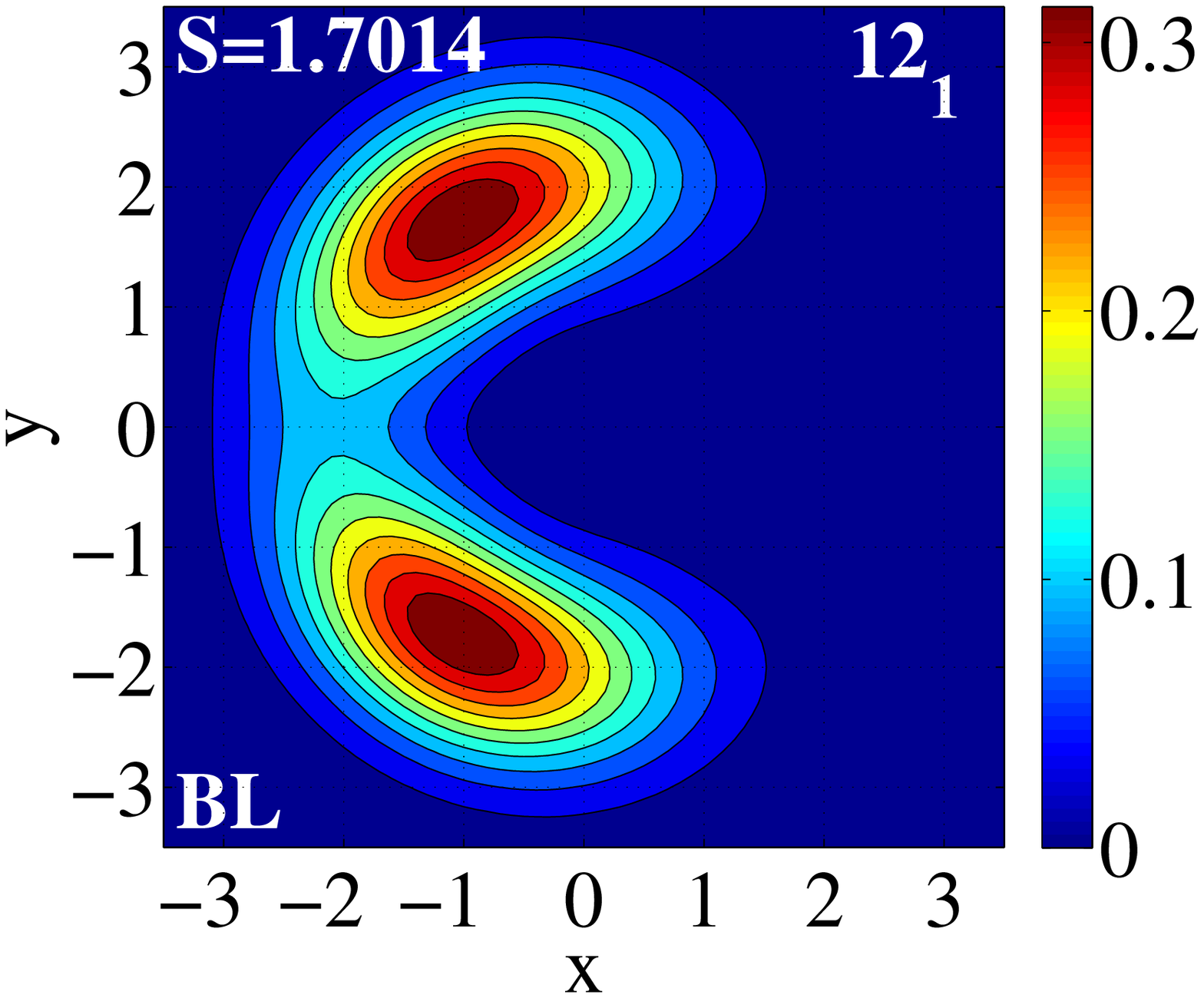}\label{fig:cpdn3l12.sub1}}
\subfigure[$~{12}_{2}$]{\includegraphics[width=0.19\linewidth]{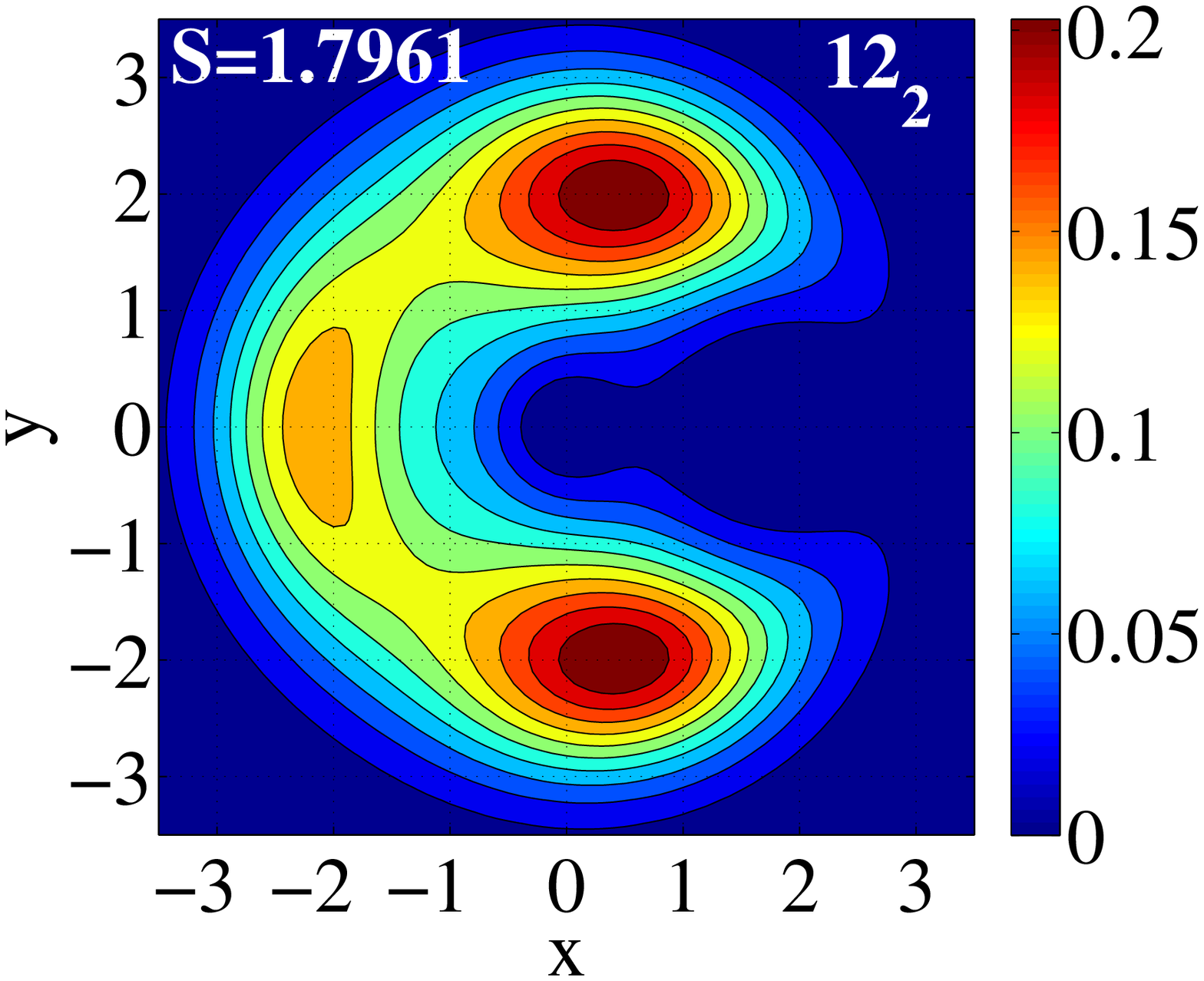}\label{fig:cpdn3l12.sub2}}
\subfigure[$~{12}_{3}$]{\includegraphics[width=0.19\linewidth]{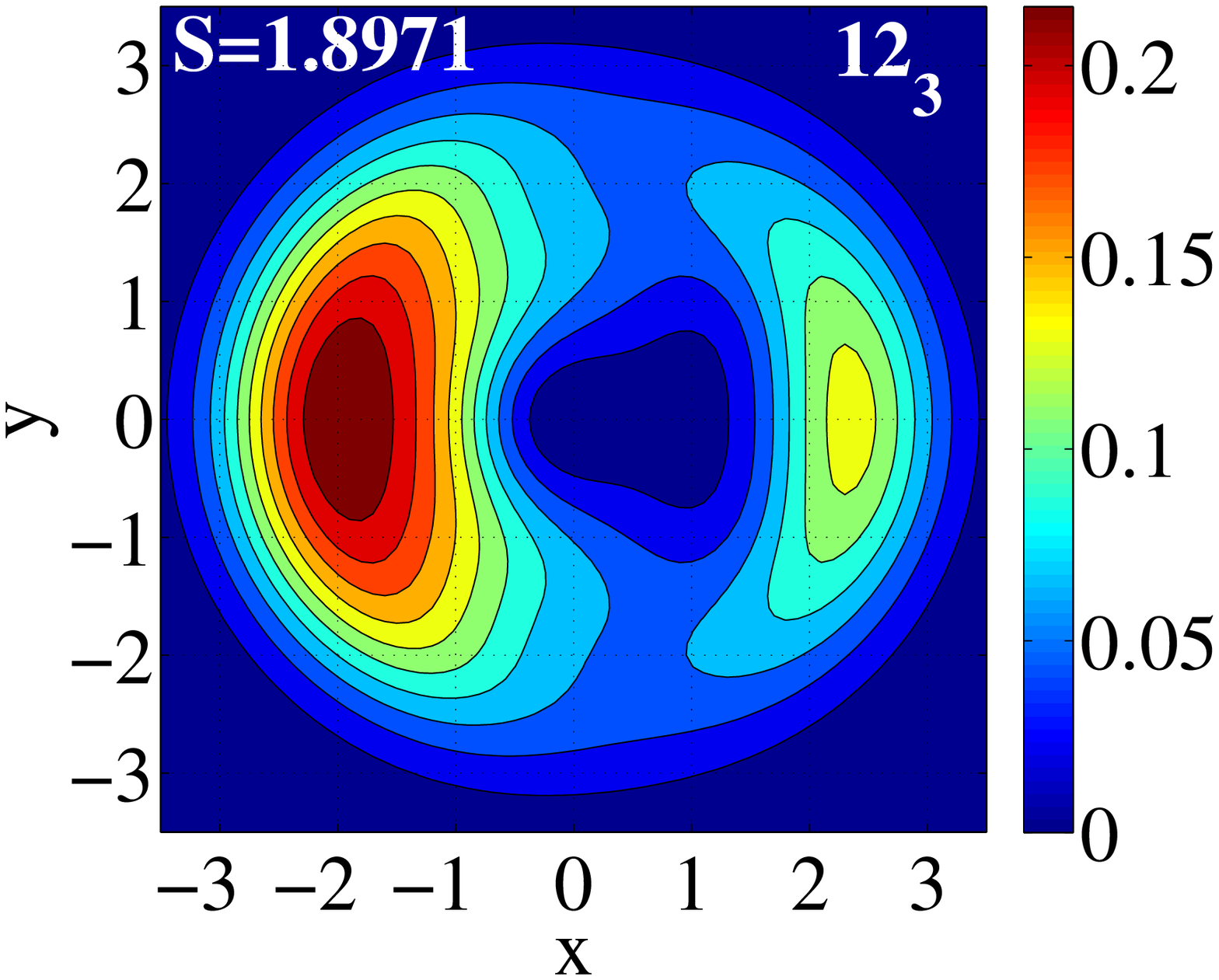}\label{fig:cpdn3l12.sub3}}
\subfigure[$~{12}_{4}$]{\includegraphics[width=0.19\linewidth]{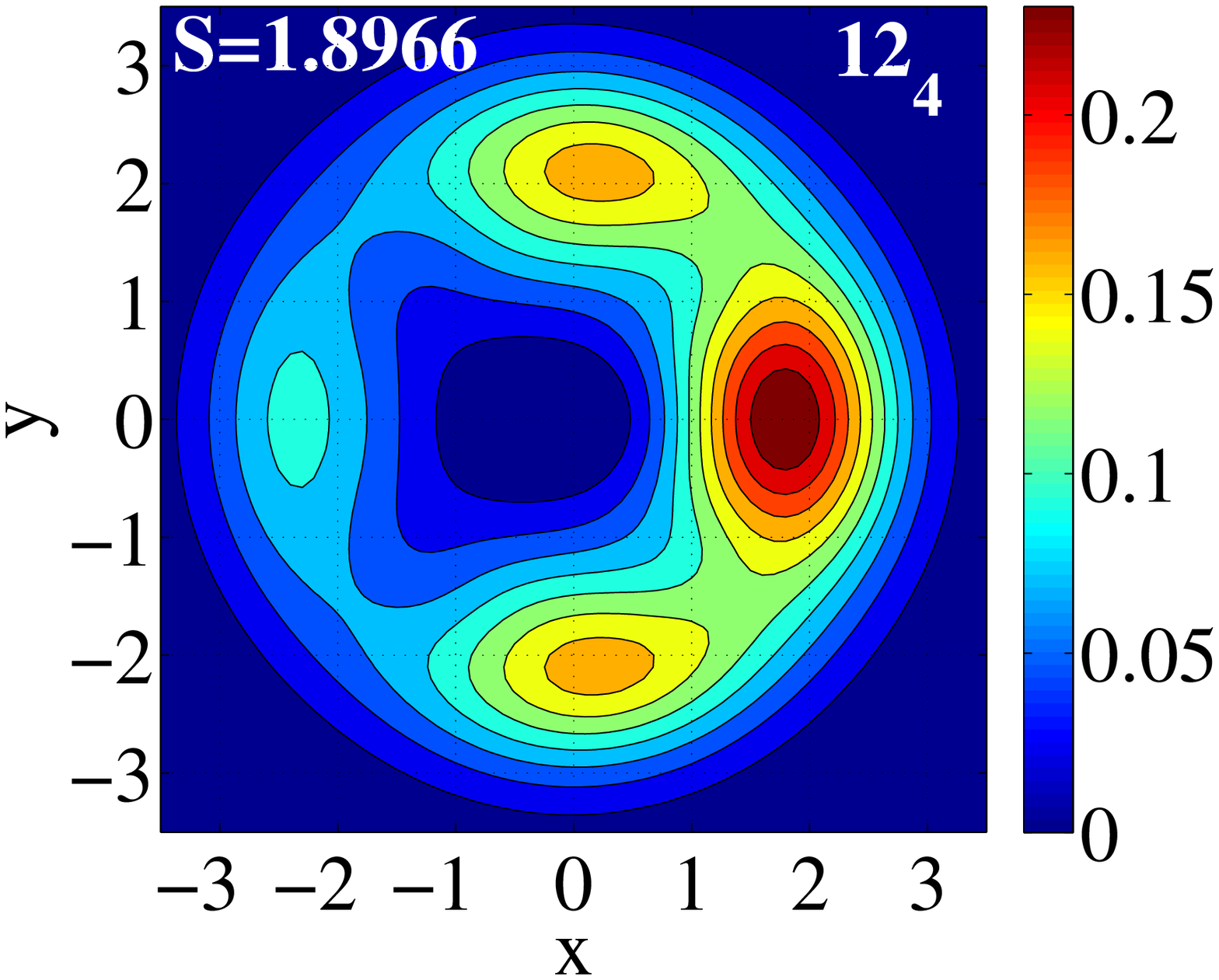}\label{fig:cpdn3l12.sub4}}
\subfigure[$~{12}_{5}$]{\includegraphics[width=0.19\linewidth]{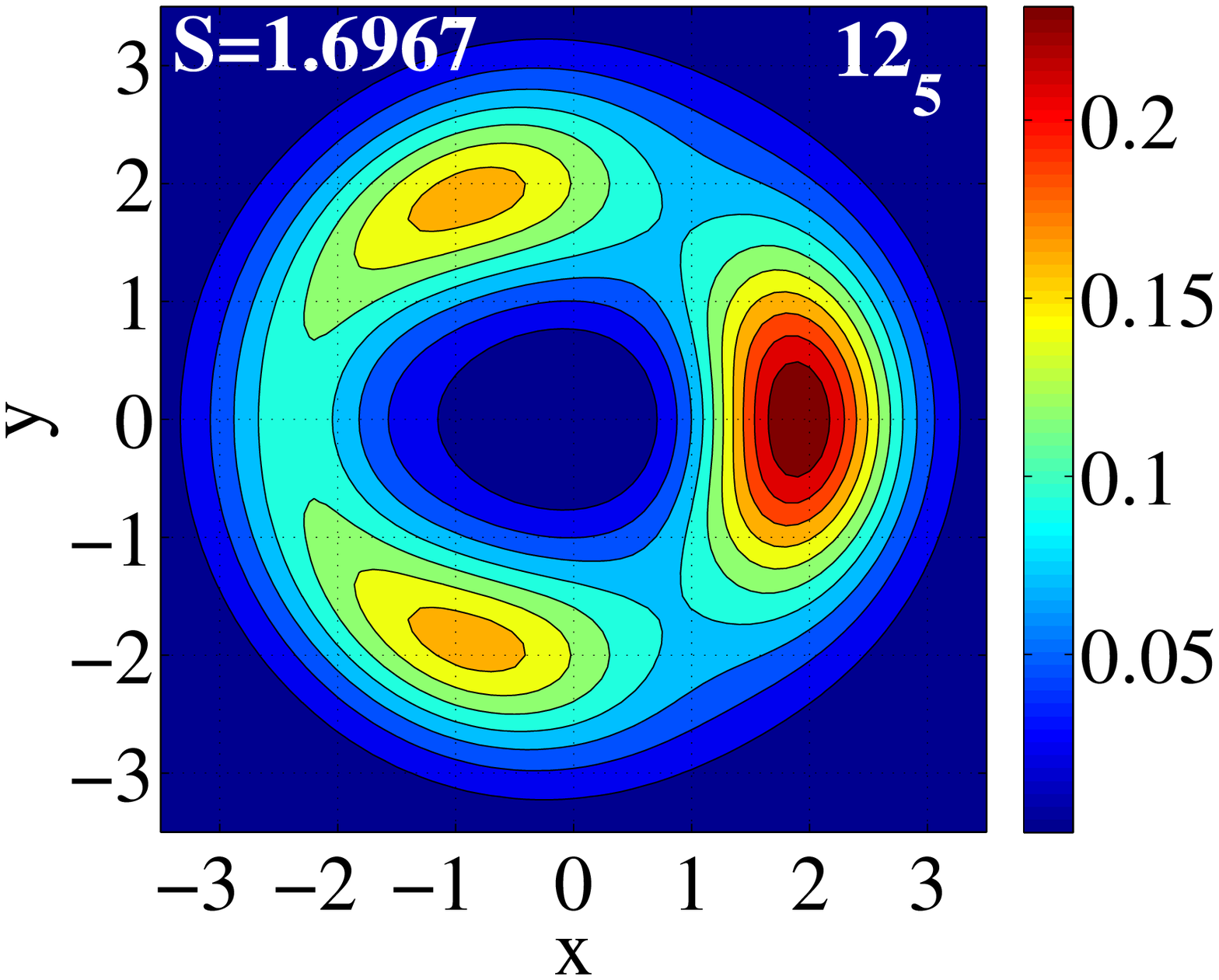}\label{fig:cpdn3l12.sub5}}
\subfigure[$~{12}_{6}$]{\includegraphics[width=0.19\linewidth]{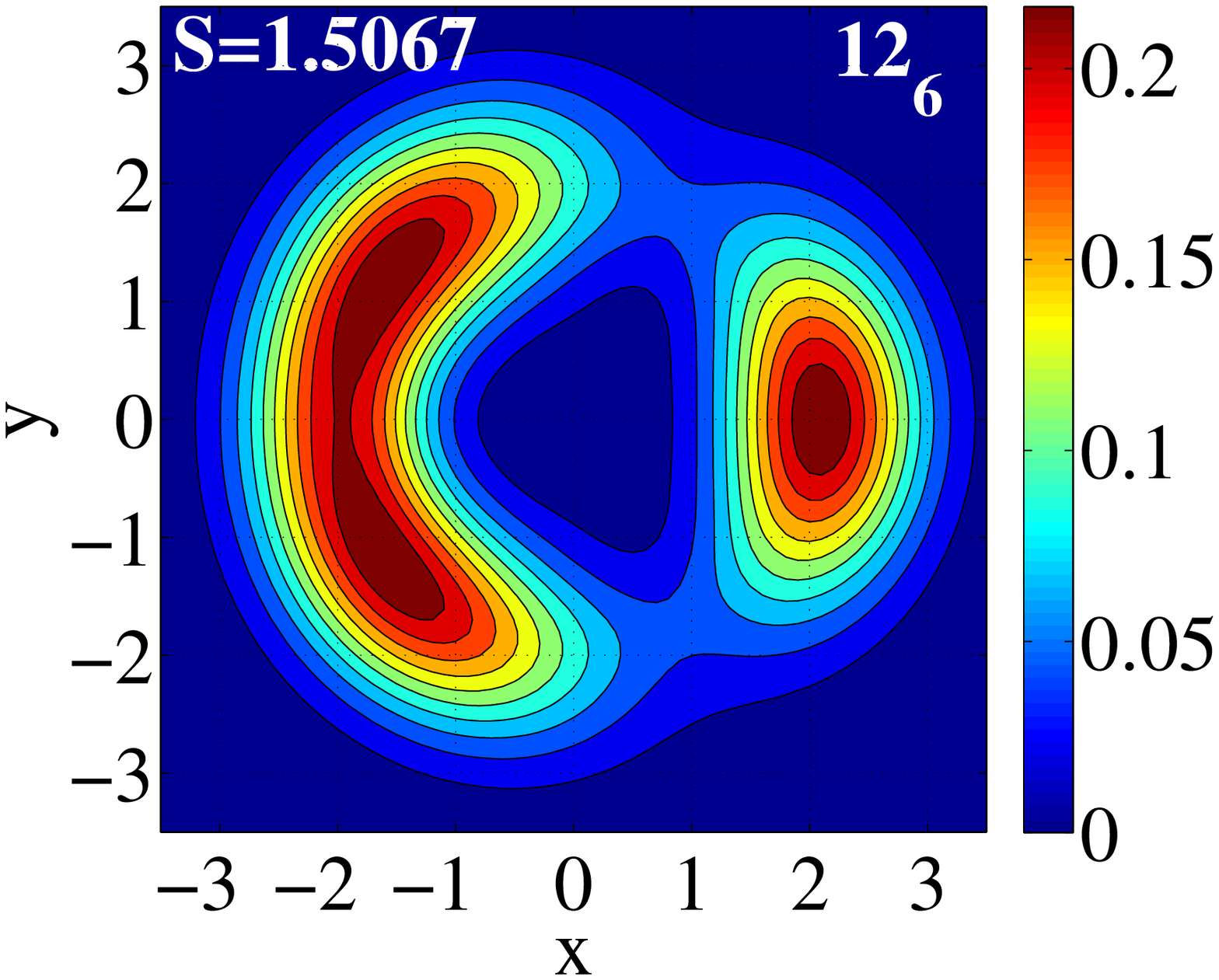}\label{fig:cpdn3l12.sub6}}
\subfigure[$~{12}_{7}$]{\includegraphics[width=0.19\linewidth]{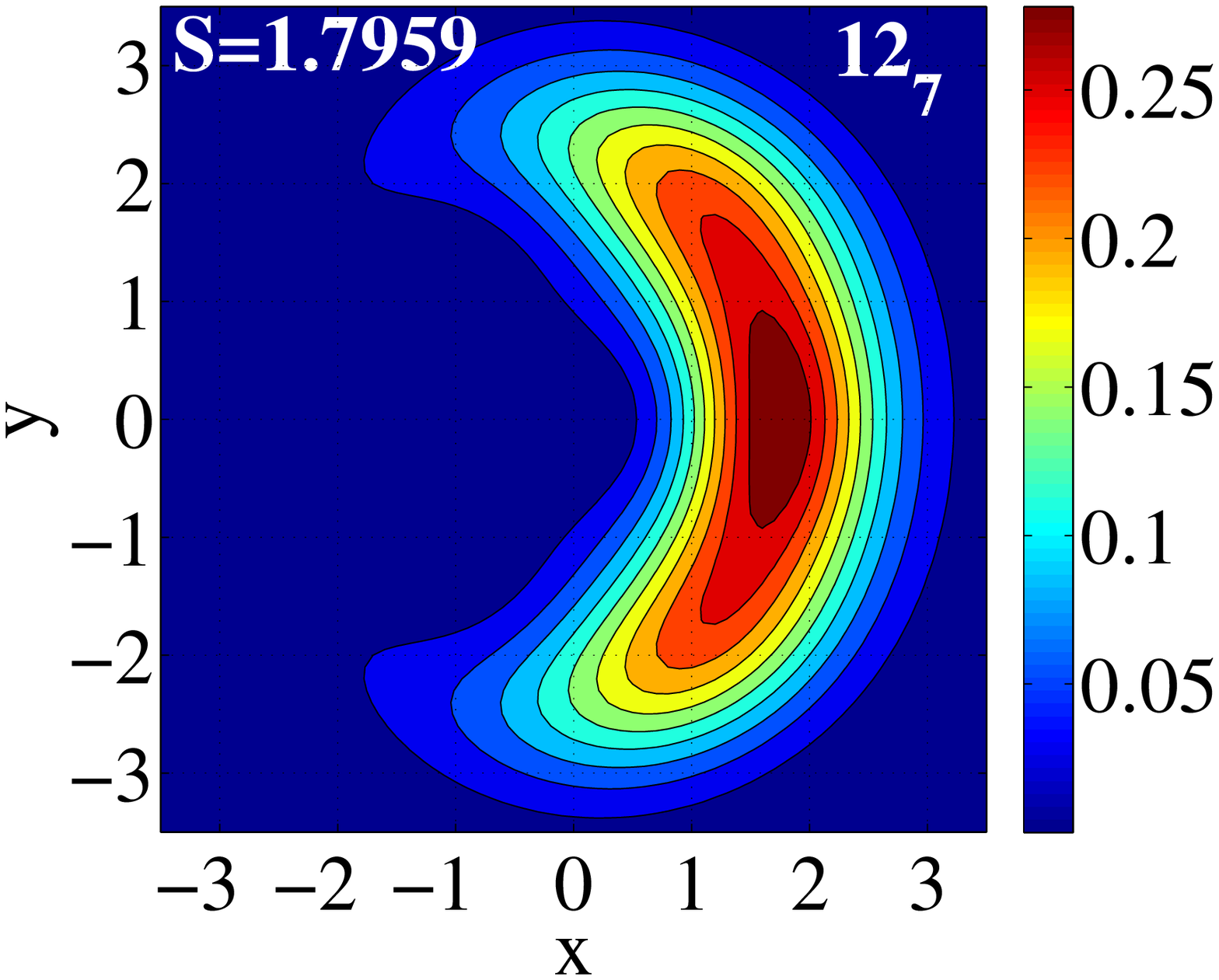}\label{fig:cpdn3l12.sub7}}
\subfigure[$~{12}_{8}$]{\includegraphics[width=0.19\linewidth]{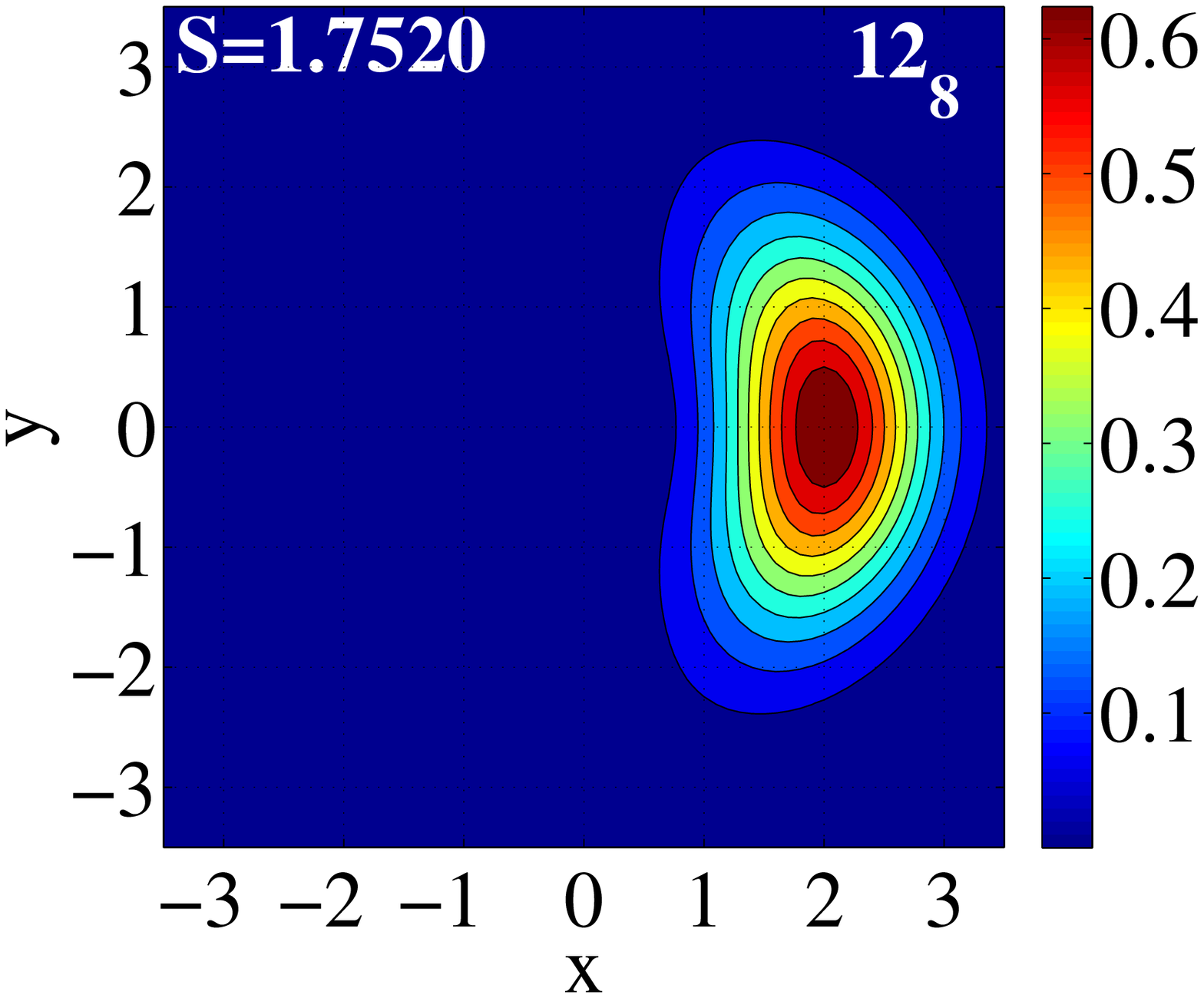}\label{fig:cpdn3l12.sub8}}
\subfigure[$~{12}_{9}$ BM]{\includegraphics[width=0.19\linewidth]{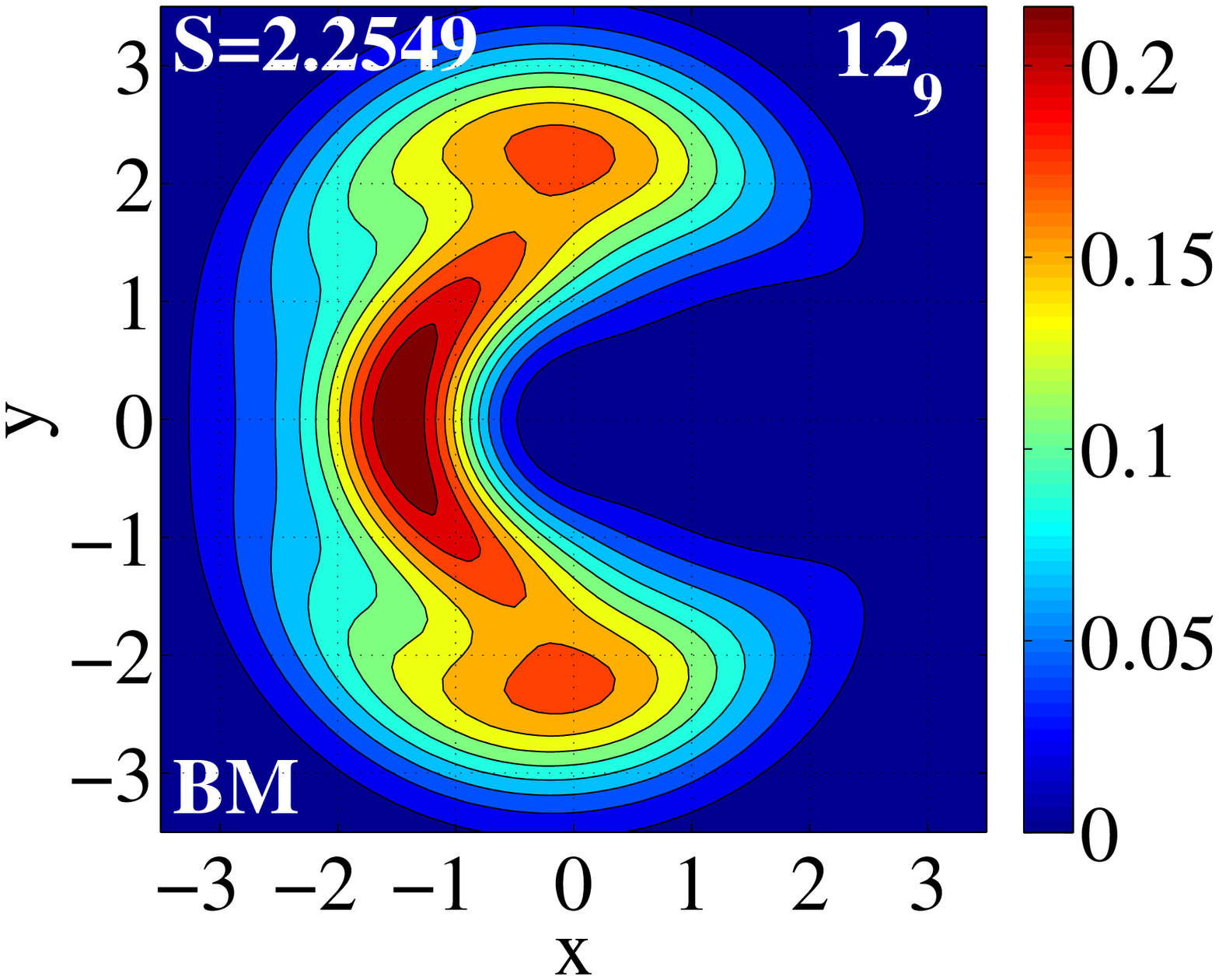}\label{fig:cpdn3l12.sub9}}
\subfigure[$~{12}_{10}$]{\includegraphics[width=0.19\linewidth]{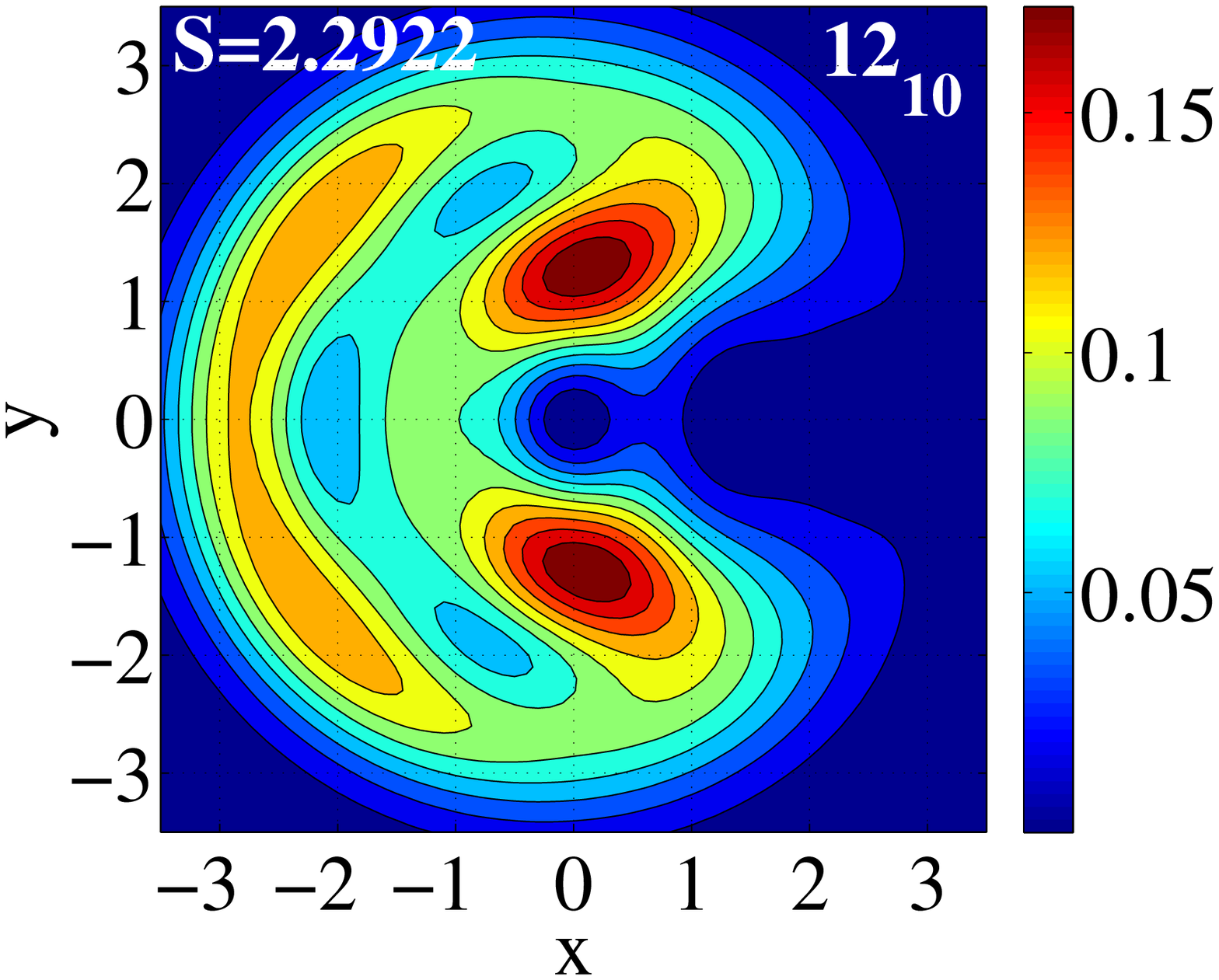}\label{fig:cpdn3l12.sub10}}
\caption{\label{fig:cpdn3l12}(Color online) Contour plots for conditional probability distribution (CPD) of various low-lying eigenstates in $L=12$ angular momentum subspace for $N=3$ bosons with $\mbox{g}_{2}=0.09151$ and $\sigma_{\perp}=0.1$ in Eq.~(\ref{gip}). The reference point ${\bf r}_{0}$ is located at $(x_{0},y_{0}) = (\sqrt{N}, 0)$ in units of $a_{\perp}$. Contour plots depict the isosurface density profiles viewed along the $z$-axis, where brown-red regions have the highest probability (of finding a particle) falling off to blue regions of low probability.}
\end{figure*}
\\
\indent 
In order to compare the internal structure of the low-lying eigenstates in $L=6$ and $L=12$ subspaces at a given value of interaction parameter $\mbox{g}_{2}=0.09151$, we further analyze CPD plots in Fig.~\ref{fig:cpdn3l6} and Fig.~\ref{fig:cpdn3l12}.
We observe that CPD profiles in Fig.~\ref{fig:cpdn3l6.sub1} and Fig.~\ref{fig:cpdn3l12.sub1} of the anticorrelated $q=2$ and $q=4$ Bose-Laughlin states $6_{1}$ and ${12}_{1}$, respectively, are similar in having two equal peaks equidistant from the reference point ${\bf r}_{0}=(1.732,0)$. 
It is also observed that the breathing modes $6_{7}$ and $12_{9}$ in Fig.~\ref{fig:cpdn3l6.sub7} and Fig.~\ref{fig:cpdn3l12.sub9} (where the system is assumed to be in the state of expansion) have two equal peaks at the opposite ends symmetrically placed with respect to the third (higher) peak as well as the reference point ${\bf r}_{0}$. 
We further observe that the states ${6}_{6}$ and ${12}_{8}$ in Fig.~\ref{fig:cpdn3l6.sub6} and Fig.~\ref{fig:cpdn3l12.sub8}, respectively, preceding the breathing modes $6_{7}$ and ${12}_{9}$ in the respective eigenspectrum, possess similar internal structures. 
The only peak in their respective CPDs is found coinciding with the reference point ${\bf r}_{0}$ implying that all the bosons are in the same position and the system is in a state of contraction. 
The close analogy between the internal structures of the low-lying eigenstates in $L=6$ and $L=12$ subspaces suggests that the (radially contracted) states $6_{6}$ and ${12}_{8}$, appear to play an important role in the emergence of the (radially expanded) first breathing modes $6_{7}$ and $12_{9}$ in respective $L$ subspaces.
\paragraph*{Effect of interaction range.} 
Through our exact diagonalization study on $N=3$ bosons, we found that the first breathing mode in $L=6$ subspace is independent of the interaction strength $\mbox{g}_{2}$ for the interaction range $\sigma_{\perp}=0.1$ in the Gaussian interaction potential~(\ref{gip}).
We, now, fix the value of interaction strength $\mbox{g}_{2}=0.09151$ and examine the effect of $\sigma_{\perp}$ on the first breathing mode in $L=6$ subspace as the interaction range is varied over $0.1 \le\sigma_{\perp}\le 0.5$.
\begin{table*}[!htb]
\caption{\label{tab:n3bm} For $N=3$ rapidly rotating bosons in total angular momentum subspace $L=6$, the values of eigenenergy ($E$) and von Neumann entropy ($S$) of the ground state and the low-lying excited states including the first breathing mode, with fixed value of interaction parameter $\mbox{g}_{2}=0.09151$ and the interaction range varied over $0.1 \le \sigma_{\perp} \le 0.5$ of the Gaussian potential~(\ref{gip}). The states $6_{1}$ and $6_{7}$ correspond to the yrast state and the first breathing mode, respectively, in $L=6$ subspace.}
\begin{ruledtabular}
\begin{tabular}{ccccccccccc}	
 & \multicolumn{2}{c}{$\sigma_{\perp}=0.1$} & \multicolumn{2}{c}{$\sigma_{\perp}=0.2$} & \multicolumn{2}{c}{$\sigma_{\perp}=0.3$} & \multicolumn{2}{c}{$\sigma_{\perp}=0.4$} & \multicolumn{2}{c}{$\sigma_{\perp}=0.5$}\\ 
\cline{2-3}\cline{4-5}\cline{6-7}\cline{8-9}\cline{10-11}\noalign{\smallskip} 
$i$ & $E(6_{i})$ & $S(6_{i})$ & $E(6_{i})$ & $S(6_{i})$ & $E(6_{i})$ & $S(6_{i})$ & $E(6_{i})$ & $S(6_{i})$ & $E(6_{i})$ & $S(6_{i})$ \\ \hline 
{\blue 1} & {\blue 13.2426} & {\blue 1.5570} & {\blue 13.2426} & {\blue 1.5569} & {\blue 13.2427} & {\blue 1.5565} & {\blue 13.2429} & {\blue 1.5557} & {\blue 13.2432} & {\blue 1.5544}\\
2 & 13.2503 & 1.6316 & 13.2501 & 1.6317 & 13.2499 & 1.6320 & 13.2496 & 1.6326 & 13.2494 & 1.6322\\
3 & 13.2522 & 1.2714 & 13.2518 & 1.2723 & 13.2515 & 1.2768 & 13.2510 & 1.2903 & 13.2505 & 1.3227\\
4 & 13.2528 & 1.5289 & 13.2525 & 1.5284 & 13.2521 & 1.5255 & 13.2516 & 1.5193 & 13.2510 & 1.5126\\
5 & 13.2566 & 1.6608 & 13.2562 & 1.6608 & 13.2556 & 1.6612 & 13.2549 & 1.6620 & 13.2543 & 1.6634\\
6 & 13.2714 & 1.5303 & 13.2706 & 1.5302 & 13.2693 & 1.5301 & 13.2677 & 1.5298 & 13.2659 & 1.5295\\
{\blue 7} & {\blue 15.2426} & {\blue 2.3204} & {\blue 15.2426} & {\blue 2.3202} & {\blue 15.2427} & {\blue 2.3193} & {\blue 15.2430} & {\blue 2.3175} & {\blue 15.2433} & {\blue 2.3143}\\
8 & 15.2451 & 2.2636 & 15.2450 & 2.2698 & 15.2448 & 2.2761 & 15.2447 & 2.2798 & 15.2447 & 2.2799\\
9 & 15.2498 & 2.1234 & 15.2494 & 2.1279 & 15.2488 & 2.1492 & 15.2480 & 2.1858 & 15.2439 & 2.2156\\
10 & 15.2512 & 2.2784 & 15.2508 & 2.3039 & 15.2504 & 2.1849 & 15.2497 & 2.2087 & 15.2491 & 2.2136\\
\end{tabular}
\end{ruledtabular}
\end{table*}
The results obtained for the ground and low-lying excited states including the first breathing mode are presented in Table~{\ref{tab:n3bm}}.  
We observe that the energy as well as the von Neumann entropy obtained for the ground state $6_{1}$ and the first breathing mode $6_{7}$ remains unchanged for the interaction range $\sigma_{\perp} < 0.3$, whereas, beyond $\sigma_{\perp} = 0.3$, both the energy and the corresponding von Neumann entropy vary with $\sigma_{\perp}$. 
Thus as $\sigma_{\perp}$ is increased and the Gaussian interaction potential deviates significantly from the $\delta$-function potential, the energy of the first breathing mode exhibits deviation from the value $2 \hbar \omega_{\perp}$ for a strictly 2D zero-range interaction potential. 
Such a study has also been presented in Ref.~\cite{iasl15} for $N=10$ bosons in $L=0$ non-rotating state. 
Our results, thus, indicate that the first breathing mode remains independent of interaction strength for small values of interaction range $\sigma_{\perp} \left(<0.3\right)$ of the Gaussian potential (\ref{gip}).
\\
\indent
The independence of the energy of the ground state and the first breathing mode for the interaction range $\sigma_{\perp} < 0.3$ of the Gaussian potential~(\ref{gip}) in $L=6$ and $L=12$ subspaces for $N=3$, can be understood in terms of the proposed variational ansatz in Eq.~(\ref{cf}) for the ground state and the first breathing mode. However, for  $\sigma_{\perp}>0.3$, the above variational ansatz breaks down.
\section{Summary and Conclusion}
\label{sec:conc}
In conclusion, we have examined the quantum correlation (von Neumann entanglement entropy) and the internal structure (CPD) of the ground and low-lying excited states of a rotating system of three bosons interacting via repulsive finite-range Gaussian potential in a quasi-2D harmonic trap.
The Hamiltonian matrix is diagonalized for given subspaces of quantized total angular momenta $0 \le L \le 4N$ in weakly to moderately interacting regime to obtain the low-lying eigenstates that provide an insight into the evolution of few-body states {\it versus} interaction and rotation.   
Our numerical results support the supposition that breathing modes, known to exist in a purely 2D system with zero-range ($\delta$-function) interaction potential, are also observed in more realistic quasi-2D system with finite-range Gaussian interaction potential. 
\\
\indent
In the rapidly rotating regime, the evolution of the Bose-Laughlin ground state and the low-lying excited states with interaction is studied, by fixing the total angular momenta $L=\frac{q}{2}N(N-1)=6$ and $L=12$ corresponding to $q=2$ and $q=4$ respectively, for the three-boson system.  
The Bose-Laughlin state which is an eigenfunction of the (zero-range) interaction potential with eigenvalue zero is strongly correlated and is indeed found to have the anticorrelation (exclusion) structure. 
We find that for the three-boson system studied here, the Bose-Laughlin state and the first breathing mode exhibit similar features such as the interaction independence of eigenenergy, von Neumann entropy and internal structure when the interaction parameter is varied over three orders of magnitude.
On the other hand, the eigenstates lying between the Bose-Laughlin state and the first breathing mode, these quantities, namely, the eigenenergy, the von Neumann entropy and the internal structure are found to vary with interaction.
\\
\indent
Moreover, the eigenstates preceding the first breathing mode in $L=6$ and $L=12$ subspaces have similar internal structures. 
The only peak in their respective CPDs is found coinciding with the position of the reference particle (so that the three bosons in the system are at the same position). 
This feature appear to play an important role in the emergence of the first breathing mode (or the breathing band). 
This is in contrast to the anticorrelated Bose-Laughlin state where the probability of finding two or more bosons at the same position is zero.
The results obtained indicate that the first breathing mode remains independent of the interaction strength for small values of the interaction range $\sigma_{\perp} \left(<0.3\right)$ of the Gaussian potential. 
For $\sigma_{\perp}>0.3$, the energy as well as the von Neumann entropy of both the ground state and the first breathing mode vary with $\sigma_{\perp}$.
Our study demonstrates that the von Neumann entropy and the CPD are powerful theoretical tools to gain insight into strongly correlated quantum states. 
We, however, wish to point out that some of the results presented here may be the consequence of special geometries like equilateral triangle, linear configuration {\it etc.}, that a three boson system can have.

\end{document}